\documentclass[12pt]{article}
\usepackage{amsmath}
\usepackage{amssymb}
\usepackage{graphicx}
\numberwithin{equation}{section}
\begin{document}

\setcounter{page}{0}
\thispagestyle{empty}

\begin{center}
{\large \bf{ Dynamical Gap Generation in Topological Insulators}}
\end{center}

\vspace*{2cm}

\begin{center}
{
Paolo Cea~\protect\footnote{Electronic address:
{\tt Paolo.Cea@ba.infn.it}}  \\[0.5cm]
{\em Dipartimento di Fisica, Universit\`a di Bari,
Bari, Italy}\\[0.3cm]
{\em INFN - Sezione di Bari, Bari,
Italy} }
\end{center}

\vspace*{1.5cm}

\renewcommand{\abstractname}{\normalsize ABSTRACT}
\begin{abstract}
We developed  a quantum field theoretical description for the surface states of three-dimensional topological insulators.
Within the relativistic quantum field theory formulation, we investigated the dynamics of low-lying surface states in an applied
transverse magnetic field. We argued that, by taking into account quantum fluctuations,  in three-dimensional topological insulators   
there is  dynamical generation of a gap by  a rearrangement of the Dirac sea. By comparing with available experimental
data we found that our theoretical results allowed a consistent and coherent description of the Landau level spectrum of
the surface low-lying excitations. Finally, we showed that the recently detected zero-Hall plateau at the charge neutral point
could be accounted for by chiral edge states  residing at the magnetic domain boundaries between the top and bottom surfaces
of the three-dimensional topological insulator. 
\end{abstract}

\vspace*{0.5cm}
\begin{flushleft}
\end{flushleft}
\newpage
\section{Introduction}
\label{S1}
Topological insulators realize  new quantum states which have recently attracted considerably
interest in condensed matter physics  (for recent reviews,  see Refs.~\cite{Hasan:2010,Moore:2010,Hasan:2011,Qi:2011,Fruchart:2013,
Ando:2013,Vafek:2014,Wehling:2014,Hasan:2014a, Hasan:2014b}  and references therein).
Topological insulators can be realized in both two dimensions  and three dimensions. In particular,
topological insulators  in three dimensions~\cite{Fu:2007}
 are non magnetic materials having an energy band gap in the bulk and possessing  metallic surface states
due to the nontrivial topology of the bulk electronic wavefunctions.
In this paper we focus  on time-reversal  invariant systems, where the nontrivial
topology is protected by time-reversal symmetry. 
In fact, in three-dimensional topological insulators the electron spin is locked to the momentum  due to time-reversal symmetry
leading to the notion of helical surface states.  
In these systems  it turns out that  the surface states display Dirac dispersions. Thus, the physics of
low-lying excitations are described by relativistic Dirac fermions.
As a result, the surface electronic band structure is similar to that of graphene 
(see, for instance, Refs.~\cite{Geim:2007,Katsnelson:2007,CastroNeto:2009,Kotov:2012}),
 except that there is just a single Dirac fermion instead of four as in graphene (due to valley and spin degeneracies).
Apparently it seems that the description of the two-dimensional
surface of  three-dimensional topological insulators in terms of a single Dirac fermion violates the Nielsen-Ninomiya no-go theorem~\cite{Nielsen:1983}
which  states that for a time-reversal invariant system Dirac points must always come in pairs. 
In fact,  the partner Dirac fermion resides on the opposite surface of the solid, restoring the counting scheme for Dirac fermions.
Finally, the surface states of a topological insulator cannot be localized even for strong
disorder as long as the bulk band gap remains intact. As a result, the surface
states are topological protect  against disorder and non-magnetic impurities.
Remarkably, the existence of these surface states has been recently confirmed by angle-resolved photoemission spectroscopy  
and scanning tunneling microscopy~\cite{Hasan:2010,Hasan:2011,Qi:2011,Ando:2013,Vafek:2014}.  \\
The low-energy excitations in three-dimensional topological insulators 
are described by an effective Hamiltonian made of a two-component Pauli spinors  which satisfy the massless
 two-dimensional Dirac equation with the speed of light replaced by the Fermi velocity $v_F$.  
If the topological insulator is   immersed in a transverse magnetic field, the relativistic massless dispersion of the electronic wave 
functions results in  non-equidistant Landau levels (cgs units):
\begin{equation}
\label{1.1}
\varepsilon^{(\pm)}_n \; =  \; E_D \; \pm \; \sqrt{2 \;  n \;  \hslash \; \frac{ v_F^2}{c} \;  eB}  \;  \;  , \;  \;  n \;  =  \;  0 \, , \,  1 \, , \,  2  \, , \;  . \, . \, .
\end{equation}
where $eB>0$, $e$ being the elementary charge, and $E_D$ is the energy of the 
charge-neutral Dirac point. Note that, here and in the following, 
we shall  not distinguish  between the magnetic flux density B and
the magnetic field H  since in a nonmagnetic material  B = H. \\
We see that the Landau quantization of massless Dirac fermions is characterized by the occurrence 
of Landau levels, called zero modes,  pinned at the Dirac point with energy $ \varepsilon_0=E_D$
and the appearance of states with energies scaling with
$\sqrt{n}$ on both the positive and negative energy sides of the Dirac point. 
In fact, the presence of anomalous Landau levels at  the Dirac point leads to the half-integer quantum Hall 
effect~\cite{Geim:2007}. Remarkably, quite recently such peculiar Landau quantization of the surface states
in three-dimensional topological insulators has been confirmed by means of the scanning tunneling
spectroscopy.  \\
The plan of the paper is as follows. In Sect.~\ref{S2} describe the dynamics of the low-lying surface states by 
means of an effective relativistic quantum field theory. In Sect.~\ref{S3} we discuss the quantum dynamics of
the surface states in presence of a transverse magnetic field.   Sect.~\ref{S4} is devoted to the discussion
of the  dynamical generation of a mass gap  by a rearrangement of the Dirac sea.
In Sect.~\ref{S5} we compare our theoretical results with available experimental data.
In Sect.~\ref{S6} we briefly discuss the quantum Hall effect, and suggest that chiral edge states could explain the 
zero-Hall plateau recently detected in three-dimensional topological insulators.
Finally, our conclusions are relegated in  Sect.~\ref{S7}. For reader's convenience some technical details are
collected  in Appendix~\ref{Ap}.
\section{Quantum Field Theory of Topological Surface States}
\label{S2}
In this section we are interested in the surface states of three-dimensional topological insulators.
More formally,  we shall describe the dynamics of these surface states by means of an effective
relativistic quantum field theory, namely quantum electrodynamics in two spatial dimensions. It turns out that 
the field theory description of such states should be adequate for our purposes. 
We, further, assume that the chemical potential is pinned at the charge-neutral Dirac point. Without loss
in generality we may set $\mu=E_D=0$. Thus, the low-lying excitations are described by the following
effective Hamiltonian~\cite{Hasan:2010,Moore:2010,Hasan:2011,Qi:2011,Ando:2013,Vafek:2014}:
\begin{equation}
\label{2.1}
\hat{H}  \; = \; \int d^2 x  \; \hat{\Psi}^{\dag}(\vec{x},t)  \left \{   - i  \; \hslash \; v_F \; \vec{\alpha} \cdot \vec{\nabla}  \right  \}  \hat{\Psi}(\vec{x},t)  \; ,
\end{equation}
where  $\vec {\alpha} = \gamma^0 \vec {\gamma}$ and $v_F$ is the Fermi velocity.
Here $ \gamma^{\mu} = (\gamma^{0},  \gamma^{1},  \gamma^{2} )$
are the Dirac  gamma matrices which satisfy the Clifford algebra :
\begin{equation}
\label{2.2}
\{\gamma^{\mu}, \gamma^{\nu}\} \; = \; 2 \; g^{\mu \nu}    \; ,
\end{equation}
$g^{\mu \nu} = diag(1,-1,-1)$ being  the  Minkowski metric  tensor. 
In two spatial dimensions a  spinor representation is provided by two-component 
Dirac spinors. Then, the fundamental representation of the Clifford algebra is given 
by $2 \times 2$ matrices which can be constructed from the Pauli matrices as follows:
\begin{equation}
\label{2.3}
\gamma^0=\sigma_3, \; \; \gamma^1 =i \sigma_1, \; \; \gamma^2=i \sigma_2 \; .
\end{equation}
The Heisenberg equations of motion are:
\begin{equation}
\label{2.4}
i \; \hslash \; \frac{\partial  \hat{\Psi} (\vec{x}, t) }{\partial t} \; = \;  \left [   \hat{\Psi} (\vec{x}, t) \, , \, \hat{H}  \right ]   \; ,
\end{equation}
\begin{equation}
\label{2.5}
i \; \hslash \; \frac{\partial  \hat{\Psi}^{\dagger} (\vec{x}, t) }{\partial t} \; = \;  \left [   \hat{\Psi}^{\dagger} (\vec{x}, t) \,  , \, \hat{H} \right ]   \; .
\end{equation}
From the equations of motion together with the canonical equal time fermion anticommutation relations:
\begin{equation}
\label{2.6}
\left \{ \hat{\Psi}(\vec{x}, t) \, , \,  \hat{\Psi}^{\dagger} (\vec{x'}, t) \right \}  = \;  \delta(\vec{x} - \vec{x'})  \; \; ,  \;  \;
\left \{ \hat{\Psi}(\vec{x}, t) \, , \,  \hat{\Psi}(\vec{x'}, t) \right \}  = 
\left \{ \hat{\Psi}^{\dagger}(\vec{x}, t) \, , \,  \hat{\Psi}^{\dagger} (\vec{x'}, t) \right \} \; = \; 0 \; ,
\end{equation}
one can determines the quantum dynamics of the system.  In fact,  we write the field operators in terms of 
the creation and annihilation operators (see, for instance, the classic textbook  Ref.~\cite{Bjorken:1968}):
\begin{equation}
\label{2.7}
 \hat{\Psi}(\vec{x}, t) \; = \;  \int d^2 p \left \{  e^{-i \frac{\varepsilon_{\vec{p}} \; t}{\hslash} }  \; 
 \psi^{(+)}(\vec{x}) \; \hat{b}_{\vec{p}}   \;  + \;   e^{+ i \frac{\varepsilon_{\vec{p}} \; t}{\hslash} }  \; 
 \psi^{(-)}(\vec{x}) \; \hat{d}^{\dagger}_{\vec{p}}  \right \}  \; ,
\end{equation}
\begin{equation}
\label{2.8}
 \hat{\Psi}^{\dagger}(\vec{x}, t) \; = \;  \int d^2 p \left \{  e^{+ i \frac{\varepsilon_{\vec{p}} \; t}{\hslash} }  \; 
 [\psi^{(+)}(\vec{x})]^{\dagger} \; \hat{b}^{\dagger}_{\vec{p}}   \;  + \;   e^{- i \frac{\varepsilon_{\vec{p}} \; t}{\hslash} }  \; 
 [\psi^{(-)}(\vec{x})]^{\dagger} \; \hat{d}_{\vec{p}}  \right \}  \; ,
\end{equation}
where $ \psi^{(\pm)}(\vec{x})$ are the positive and negative energy solutions of the well known Dirac equation (see Appendix~\ref{Ap})
 and the  creation and annihilation operators satisfy the anticommutation  relations:
\begin{equation}
\label{2.9}
\left \{ \hat{b}^{\dagger}_{\vec{p}}  \, , \,  \hat{b}_{\vec{p'}} \right \} \; = \; \delta(\vec{p} - \vec{p'})  \; \; , \; \; 
\left \{ \hat{d}^{\dagger}_{\vec{p}}  \, , \,  \hat{d}_{\vec{p'}} \right \} \; = \; \delta(\vec{p} - \vec{p'})  
\end{equation}
with all the other anticommutators vanishing.  The  $\hat{b}^{\dagger}_{\vec{p}} $ and  $\hat{b}_{\vec{p}}$ operators create and destroy
particle above the Fermi sea, while  $\hat{d}^{\dagger}_{\vec{p}}$ and $ \hat{d}_{\vec{p}}$ operators create and destroy hole inside the
Fermi sea. 
We rewrite the Hamiltonian in terms of the  the creation and annihilation operators. To this end we insert Eqs.~(\ref{2.7})  and (\ref{2.8})
into Eq.~(\ref{2.1}). Using  Eqs.~(\ref{A.4})  and (\ref{A.5}) we obtain:
\begin{equation}
\label{2.10}
\hat{H}  \; = \; \int d^2 p  \;   \varepsilon_{\vec{p}} \;  \left \{     
 \hat{b}^{\dagger}_{\vec{p}}  \;  \hat{b}_{\vec{p}} \; -  \;     \hat{d}_{\vec{p}}  \; \hat{d}^{\dagger}_{\vec{p}}  
 \right  \}    \;  =  \;   \int d^2 p  \;   \varepsilon_{\vec{p}} \;  \left \{     
 \hat{b}^{\dagger}_{\vec{p}}  \;  \hat{b}_{\vec{p}} \; +  \;    \hat{d}^{\dagger}_{\vec{p}} \;  \hat{d}_{\vec{p}}  \right  \}  \; + E_0 \; , 
\end{equation}
where:
\begin{equation}
\label{2.11}
E_0 \; =    \;  - \;   \int d^2 p  \;   \varepsilon_{\vec{p}}  \;  \delta(\vec{p} - \vec{p})  \; . 
\end{equation}
To properly interpret the $\delta$-function in   Eq.~(\ref{2.11}), we note that:
\begin{equation}
\label{2.12}
 \delta(\vec{q})  \; =    \;   \;   \int \;  \frac{d^2 x } {(2 \pi \hslash)^2}   \;    e^{+  i \;  \frac{\vec{q} \cdot \vec{x}}{\hslash} }  \; , 
\end{equation}
so that:
\begin{equation}
\label{2.13}
 \delta(\vec{q}=0)  \; =    \;   \;   \int \;  \frac{d^2 x } {(2 \pi \hslash)^2}   \; =  \; \frac{V} {(2 \pi \hslash)^2}  \; ,
\end{equation}
where $V$ is the two-dimensional volume (area) of the system. Accordingly we have:
\begin{equation}
\label{2.14}
E_0  \;  =    \;  - \; V \;   \int  \;   \frac{d^2 p }{(2 \pi \hslash)^2}   \;  \;  \varepsilon_{\vec{p}}  \;  . 
\end{equation}
The Hamiltonian  Eq.~(\ref{2.10}) is  now normal ordered with respect to filled Fermi-Dirac sea.
In fact, $E_0$ given by Eq.~(\ref{2.14}) is the energy of the ground state (the vacuum) defined by:
\begin{equation}
\label{2.15}
  \hat{b}_{\vec{p}} \; | 0 > \; = \;    \hat{d}_{\vec{p}} \; | 0 > \; = \; 0  \;  . 
\end{equation}
It is now evident from  Eq.~(\ref{2.15}) that the ground state corresponds to the filled Fermi-Dirac sea.
If we try to evaluate the vacuum energy $E_0$, we face with the problem of ultraviolet divergencies. In fact, using
$ \varepsilon_{\vec{p}}  =  v_F \; | \vec{p} |$ we obtain:
\begin{equation}
\label{2.16}
E_0  \;  =    \;  - \; V  \;   \frac{v_F }{2 \pi \hslash^2}   \int_0^{\infty}  \;  dp    \;  p^2  \;  . 
\end{equation}
As is well known, the problem of divergent quantum corrections is inherent to relativistic quantum field theories. The infinities encountered in 
relativistic quantum theories can be cured with a procedure  called renormalization~~\cite{Bjorken:1968}.  Actually,  in condensed
matter physics one never had to deal with ultraviolet divergences. In fact, the description of the low-lying excitations by means
of the effective Hamiltonian  Eq.~(\ref{2.10}) is valid as long as $  \varepsilon_{\vec{p}} \, \lesssim \, W$, where $W$ is the band-width.
Therefore there is a natural ultraviolet cut-off which makes finite the vacuum energy  Eq.~(\ref{2.16}):
\begin{equation}
\label{2.17}
E_0  \;  \simeq    \;  - \; V  \;   \frac{1}{6 \, \pi} \; \frac{W^3}{ \hslash^2\, v_F^2}    \;  . 
\end{equation}
This (negative) vacuum energy contributes to the binding energy of the whole system and, therefore, does not
influence the dynamics of the low-lying excitations. In this case the renormalization corresponds to
simply subtract $E_0$ from the Hamiltonian operator  $\hat{H}$ leading to the renormalized Hamiltonian:
\begin{equation}
\label{2.18}
\hat{H}_R  \; \equiv \;  \hat{H} \; - \; E_0 \; = \; 
 \int d^2 p  \;   \varepsilon_{\vec{p}} \;  \left \{     
 \hat{b}^{\dagger}_{\vec{p}}  \;  \hat{b}_{\vec{p}} \; +  \;    \hat{d}^{\dagger}_{\vec{p}} \;  \hat{d}_{\vec{p}}  \right  \}  \; . 
\end{equation}
This renormalization corresponds to the so-called normal ordering prescription. For the purposes of the present work
we do not have to do any further renormalization. \\
For later convenience, we rewrite the ground-state energy  as:
\begin{equation}
\label{2.19}
E_0  \;  =    \;  - \; V \;   \int  \;   \frac{d^2 p }{(2 \pi \hslash)^2}   \;  \;  \  \sqrt{v_F^2 \, \vec{p}^2}  \;  . 
\end{equation}
To evaluate the integral we use the following  computational trick which
has been has been already  employed by us~\cite{Cea:1985}:
\begin{equation}
\label{2.20}
\sqrt{a} \; = \; - \; \int_0^{\infty} \frac {d s} {\sqrt{ \pi s}} \; \frac {d} {d s} \; 
 e^{- a s}  \; .
\end{equation}
We get:
\begin{equation}
\label{2.21}
E_0   =    V    \int  \;   \frac{d^2 p }{(2 \pi \hslash)^2}   \; \int_0^{\infty} \frac {d s} {\sqrt{ \pi s}} \; \frac {d} {d s} \; 
 e^{- v_F^2 \, \vec{p}^2 \, s}   =     V     \frac{1}{4 \, \pi} \;  \frac{1}{ \hslash^2\, v_F^2}  
 \; \int_0^{\infty} \frac {d s} {\sqrt{ \pi s}} \; \frac {d} {d s} \left (   \frac {1} {s} \right ) \; .
 \end{equation}
Equation (\ref{2.21}) shows that  the ultraviolet divergency manifest itself as a singularity for $ s \to  0$. Introducing
an effective band-width $\tilde W$ such that $ s \ge  1/\tilde{W}^2$, we easily obtain:
\begin{equation}
\label{2.22}
E_0  \;  \simeq    \;  - \; V  \;   \frac{1}{3 \, \pi \sqrt{\pi}} \; \frac{\tilde{W}^3}{ \hslash^2\, v_F^2}    \;  . 
\end{equation}
Comparing this last equation with  Eq.~(\ref{2.17}) we see that the cut-off   $\tilde W$ is finitely related to the
cut-off  $W$,   $\tilde{W} = (\frac{\sqrt{\pi}}{2})^{1/3} W$.
\section{Topological  Fermions in  Magnetic Fields}
\label{S3}
A distinguish feature of two-dimensional Dirac fermions is the peculiar Landau quantization of the energy states
in applied magnetic fields.  In presence of an external magnetic field  the effective Hamiltonian operator
Eq.~(\ref{2.1}) becomes:
\begin{equation}
\label{3.1}
\hat{H}  \; = \; \int d^2 x  \; \hat{\Psi}^{\dag}(\vec{x},t)  \left \{ v_F  \;  \vec{\alpha} \cdot \left [ - i  \; \hslash \;  \vec{\nabla} 
\; + \; \frac{e}{c} \, \vec{A}(\vec{x}) \right ] \right  \}  \hat{\Psi}(\vec{x},t)  \; ,
\end{equation}
where the electromagnetic vector potential is such that $ \vec{\nabla} \times  \vec{A}(\vec{x}) = \vec{B}(\vec{x})$. 
To write the Hamiltonian operator  in terms of   the creation and annihilation operators, 
it is convenient to  expand the fermion field operator $ \hat{\Psi}^{\dag}(\vec{x},t)$ 
and  $ \hat{\Psi}(\vec{x},t)$  in terms  of the wave functions which are the solutions of the Dirac equation in presence 
of the external magnetic field. In other words, we shall adopt the so-called  Furry picture~\cite{Schweber:1961}.  Thus,
we   expand the fermion field operator $ \hat{\Psi}^{\dag}(\vec{x},t)$  and  $ \hat{\Psi}(\vec{x},t)$ in terms  of the wave function basis 
$\psi^{(+)}_{n,p}$ and  $\psi^{(-)}_{n,p}$ given in Appendix~\ref{Ap}:
\begin{equation}
\label{3.2}
 \hat{\Psi}(\vec{x}, t) =  \int^{+\infty}_{-\infty} 
d p  \left \{    \sum^{\infty}_{n=1} \left [  e^{-i \frac{\varepsilon_{n} \; t}{\hslash} }  \; 
 \psi^{(+)}_{n,p}(\vec{x}) \; \hat{b}_{n,p}  \,  + \,   e^{+ i \frac{\varepsilon_{n} \; t}{\hslash} }  \;
 \psi^{(-)}_{n,p}(\vec{x}) \; \hat{d}^{\dagger}_{n,p}   \right ] \, + \,
 \psi_{0,p}(\vec{x}) \; \hat{c}_{p} 
  \right \}  \; ,
\end{equation}
\begin{equation}
\label{3.3}
\hat{\Psi}^{\dagger}(\vec{x}, t)  =  \int^{+\infty}_{-\infty} 
d p  \left \{  \sum^{\infty}_{n=1} \left [ e^{+i \frac{\varepsilon_{n} \; t}{\hslash} }  \; 
 \psi^{(+)\dagger}_{n,p}(\vec{x}) \; \hat{b}^{\dagger}_{n,p}   \,  + \,   e^{- i \frac{\varepsilon_{n} \; t}{\hslash} }  \;
 \psi^{(-)\dagger}_{n,p}(\vec{x}) \; \hat{d}_{n,p} \right ] \, + \,
 \psi^{\dagger}_{0,p}(\vec{x}) \; \hat{c}^{\dagger}_{p} 
  \right \}  \; .
\end{equation}
The creation and annihilation operators operators satisfy the standard  anticommutation relations:
\begin{equation}
\label{3.4}
\left \{ \hat{b}^{\dagger}_{n',p'}  \, , \,  \hat{b}_{n,p} \right \}  =  \delta(p - p') \, \delta_{n,n'}  \; ,  \; 
\left \{ \hat{d}^{\dagger}_{n',p'}  \, , \,  \hat{d}_{n,p} \right \}  =  \delta(p - p') \, \delta_{n,n'}  \; , \; 
\left \{ \hat{c}^{\dagger}_{p'}  \, , \,  \hat{c}_{p} \right \}  = \delta(p - p')  \; , 
\end{equation}
all the other anticommutators vanishing. Inserting  Eqs.~(\ref{3.2})  and  (\ref{3.3}) into  the Hamiltonian operator  Eqs.~(\ref{3.1})
we find:
\begin{equation}
\label{3.5}
\hat{H}   =    \sum^{\infty}_{n=1} \int^{+\infty}_{-\infty}  d p \,   \varepsilon_{n}
 \left \{  \hat{b}^{\dagger}_{n,p}  \, \hat{b}_{n,p}   -      \hat{d}_{n,p}  \; \hat{d}^{\dagger}_{n,p}  
 \right  \}     =   
  \sum^{\infty}_{n=1} \int^{+\infty}_{-\infty}  d p  \, \varepsilon_{n}
  \left \{   \hat{b}^{\dagger}_{n,p}  \,  \hat{b}_{n,p}  +      \hat{d}^{\dagger}_{n,p} \,  \hat{d}_{n,p}  \right  \}   +  E_0(B)  \; , 
\end{equation}
with:
\begin{equation}
\label{3.6}
E_0(B) \;  =  \; - \;  \sum^{\infty}_{n=1} \int^{+\infty}_{-\infty}  d p \,   \varepsilon_{n} \;  \delta(p-p)   \, \delta_{n,n} \; 
= \; - \; V \;  \frac {eB} {2 \pi \hslash c}  \; \sum^{\infty}_{n=1}  \,   \varepsilon_{n} \; .
\end{equation}
The physical interpretation of $E_0(B) $ is straightforward. In fact, from  Eq.~(\ref{3.5}) we infer that the ground state of the Hamiltonian
is given by:
\begin{equation}
\label{3.7}
  \hat{b}_{n,p} \; | 0 ; B > \; = \;    \hat{d}_{n,p} \; | 0 ; B > \; = \; 0   \; \; ,
\end{equation}
so that:
\begin{equation}
\label{3.8}
 \hat{H} \; | 0 ; B > \; = \;   E_0(B)  \; | 0 ; B >   
\end{equation}
namely,   $E_0(B) $  is the vacuum energy in presence of the external magnetic field B.  \\
Since for large $n$ $   \varepsilon_{n}  \sim \sqrt{n}$ (see Appendix~\ref{Ap}), we see that the summation over $n$ in   Eq.~(\ref{3.6})
 is divergent. In other words,  $E_0(B) $  is affected by ultraviolet divergences. However, if we consider the renormalized Hamiltonian 
 Eq.~(\ref{2.18}) we find: 
\begin{equation}
\label{3.9}
\hat{H}_R   \; =  \;   \hat{H} \; - \; E_0 \;  = \;  \sum^{\infty}_{n=1} \int^{+\infty}_{-\infty}  d p  \, \varepsilon_{n}
  \left \{   \hat{b}^{\dagger}_{n,p}  \,  \hat{b}_{n,p}  +      \hat{d}^{\dagger}_{n,p} \,  \hat{d}_{n,p}  \right  \}   +  \Delta E_0(B)  \; , 
\end{equation}
where: 
\begin{equation}
\label{3.10}
 \Delta E_0(B)  \;  = \;   E_0(B)  \;  - \; E_0  \; .   
\end{equation}
Now we show that $ \Delta E_0(B) $ is not affected by any ultraviolet divergences. In fact, in Appendix~\ref{Ap} we show that:
\begin{equation}
\label{3.11}
 \Delta E_0(B)  \;  =  \;   E_0(B)  \, - \, E_0 \; = 
 \frac {eB} {2 \pi \hslash c}  \, V   \int_0^{\infty} \frac {ds} {\sqrt{\pi s}} \frac 
{d} {ds} \left [    \frac{1}{1 -   e^{  - 2 \hslash v_F^2  \frac{eB}{c}  s}} \;   -  \; \frac {1} {   2 \hslash v_F^2  \frac{eB}{c}  s    }
   \right ] \; . 
\end{equation}
 This last equation shows that, indeed,  $\Delta E_0(B) $ is finite. After some elementary  manipulations we rewrite  Eq.~(\ref{3.11})
as:  
\begin{equation}
\label{3.12}
 \Delta E_0(B)  \;  =   \frac {eB} {2 \pi \hslash c}  \, V   \, v_F \, \sqrt{ 2 \hslash   \frac{eB}{c} } \; \times \;  I \; ,
\end{equation}
where:
\begin{equation}
\label{3.13}
I \; =   \int_0^{\infty} \frac {dx} {\sqrt{\pi x}} \frac 
{d} {dx} \left [    \frac{1}{1 -   e^{ - x}} \;   -  \; \frac {1} {  x   }   \right ] \; = 
 \int_0^{\infty} \frac {dx} {\sqrt{\pi x}}
 \left [    \frac{e^{-x}}{(1 -   e^{ - x})^2} \;   +  \; \frac {1} {  x^2   }   \right ]  \;  \; \simeq \; 0.208 \; . 
\end{equation}
We end, thus, with the remarkable result that in presence of an applied magnetic field the ground state
 due to quantum fluctuations acquires a finite positive energy given by   Eq.~(\ref{3.12}). 
To understand the physical meaning of $\Delta E_0(B)$, we observe that the degeneracy of the Landau levels, namely the number
of states in a given level, is:
\begin{equation}
\label{3.14}
g_L \;  =   \frac {eB} {2 \pi \hslash c}  \, V  \;  \; .
\end{equation}
Therefore, we may define the vacuum energy per particle as:
\begin{equation}
\label{3.15}
 \varepsilon_{P}(B)  \; \equiv \;  \frac{\Delta E_0(B)}{g_L}   \;  \simeq \; 0.208 \;  v_F \;  \sqrt{ 2 \hslash   \frac{eB}{c} } \;  \;  .
\end{equation}
Now, it is quite easy to convince ourself that   $\varepsilon_{P}(B)$ modifies the single particle spectrum  Eq.~(\ref{1.1}) as:

\begin{equation}
\label{3.16}
\varepsilon^{(\pm)}_n \; =  \; E_D \, + \, \varepsilon_{P}(B) \; \pm \; \sqrt{2 \, \hslash \, v_F \, \frac{eB}{c} \,n}  \;  , \;  n \;  =  \;  0 \, , \,  1 \, , \,  2  \, , \; ...
\end{equation}
Usually one assumes that the external magnetic field does not modify the Dirac point energy. However, the  subtle quantum effects
we have evaluated show that  external magnetic fields give rise to a shift of the Dirac point. Remarkably, this effect manifest itself
also in graphene~\cite{Cea:2012,Kandemir:2015}.  We believe that the effect has been already observed in the experimental investigation
of the quantum Hall effect.  In fact, it turns out that  by switching on an external magnetic field one must tune the gate voltage $V_g$
to drive the system to the Dirac neutral point. Moreover, it seems that the gate voltage needed to reach the neutral point
depends on the magnetic field as $V_g \sim \sqrt{B}$, in qualitative agreement with our Eq.~(\ref{3.15}).
\section{Dynamical Generation of Mass Gap}
\label{S4}
In Sect.~\ref{S2} we said that in  three-dimensional topological insulators the low-lying excitations are described by the 
effective Hamiltonian Eq.~(\ref{2.1}). Indeed, this Hamiltonian corresponds to massless two-dimensional
Weyl fermions. In two dimensions  we can define the parity and time-reversal transformation~\cite{Jackiw:1981,Schonfeld:1981,Deser:1982}
according to:
\begin{eqnarray}
\label{4.1}
{\cal {P}}  A^0 (\vec{x}, t) {\cal {P}}^{-1} & = & A^0(\vec{x'}, t) 
\nonumber  \\
{\cal {P}}  A^1 (\vec{x}, t) {\cal {P}}^{-1} & =& - A^1(\vec{x'}, t) \\
 {\cal {P}}  A^2 (\vec{x}, t) {\cal {P}}^{-1} & =&  A^2(\vec{x'}, t)
\nonumber \\
 {\cal {P}} \; \Psi(\vec{x}, t) \; 
{\cal {P}}^{-1} &=& \sigma_1 \Psi(\vec{x'}, t) \;  \; ,
\nonumber 
\end{eqnarray}
where $\vec{x}=(x_1,x_2)$ and $\vec{x'}=(-x_1,x_2)$,
\begin{eqnarray}
\label{4.2}
{\cal {T}}  A^0 (\vec{x}, t) {\cal {T}}^{-1} & = & A^0(\vec{x}, -t) 
\nonumber  \\
{\cal {T}} \; {\vec {A}} \; (\vec{x}, t) {\cal {T}}^{-1} & =&  
-{\vec {A}} (\vec{x}, -t) \\
{\cal {T}} \; \Psi(\vec{x}, t) \; 
{\cal {T}}^{-1} &=& \sigma_2 \Psi(\vec{x}, -t) \;  \; .
\nonumber 
\end{eqnarray}
It is, now, easy to show that the Hamiltonian  Eq.~(\ref{2.1}) is invariant under  ${\cal {P}}$ and ${\cal {T}}$ transformations. 
However, if one is interested in massive fermions, then the relevant Hamiltonian turns out to be:
\begin{equation}
\label{4.3}
\hat{H}  \; = \; \int d^2 x  \; \hat{\Psi}^{\dag}(\vec{x},t)  \left \{   - i  \; \hslash \; v_F \; \vec{\alpha} \cdot \vec{\nabla} 
\; + \; m \, v_F^2 \, \sigma_3 \;  \right  \}  \hat{\Psi}(\vec{x},t)  \; .
\end{equation}
It is easy to recognize that  the mass term is  odd under both ${\cal {P}}$ and ${\cal {T}}$ transformations. It is important to point out 
that this parity-violating mass is, in fact, the only possibility for two-dimensional Weyl fermions. \\
Long time ago we showed~\cite{Cea:1985} that in (2+1)-dimensional quantum electrodynamics with two-component
Dirac fermions in an external magnetic fields it is energetically favored to develop a fermion mass gap.
In the present Section we shall show that  this happens also to   the surface states of three-dimensional topological insulators. \\
In presence of applied magnetic fields the relevant Hamiltonian is given by Eq.~(\ref{3.1}). 
Since we are interested in massive fermions, then the effective Hamiltonian becomes:
\begin{equation}
\label{4.4}
\hat{H}  \; = \; \int d^2 x  \; \hat{\Psi}^{\dag}(\vec{x},t)  \left \{ v_F  \;  \vec{\alpha} \cdot \left [ - i  \; \hslash \;  \vec{\nabla} 
\; + \; \frac{e}{c} \, \vec{A}(\vec{x}) \right ]  \; + \; m \, v_F^2 \, \sigma_3 \;    \right  \}  \hat{\Psi}(\vec{x},t)  \; .
\end{equation}
The results of  Ref.~\cite{Cea:1985} would  imply that it is energetically favored to develop a  gap such that:
\begin{equation}
\label{4.5}
 m \;  v_F^2  \; = \; \Delta_0 \; sign(B)  \; ,
\end{equation}
 with $\Delta_0  >  0 $. Actually, the magnetic field is a vector. However, we are considering magnetic fields which are perpendicular
 to the surfaces of the specimen. Thus, as concern the dynamics of the surface quasiparticles the magnetic field can be 
 considered a (pseudo)scalar. Our convention is that $B > 0$ if the magnetic field is directed along the normal to the surface, while
 $B < 0$ otherwise. Note that under parity and time-reversal transformations both $B$ and $m$ change sign, so that the gap
 $\Delta_0$ in Eq.~(\ref{4.5})  is  ${\cal {P}}$ and ${\cal {T}}$ invariant . \\
 To be definite, we fix the direction of the magnetic field such that $B > 0$. Accordingly we are left with the following Hamiltonian:
\begin{equation}
\label{4.6}
\hat{H}  \; = \; \int d^2 x  \; \hat{\Psi}^{\dag}(\vec{x},t)  \left \{ v_F  \;  \vec{\alpha} \cdot \left [ - i  \; \hslash \;  \vec{\nabla} 
\; + \; \frac{e}{c} \, \vec{A}(\vec{x}) \right ]  \; + \; \Delta_0 \, \sigma_3 \;    \right  \}  \hat{\Psi}(\vec{x},t)  \; .
\end{equation}
To write the Hamiltonian  in the  Furry representation, we need the solutions of the Dirac equation:
\begin{equation}
\label{4.7}
 \left \{  v_F \; \vec {\alpha} \cdot  \left [  -i   \hslash \vec {\nabla} \; + \;  \frac{e}{c}  \vec { A}(\vec{x})   \right ]  + \; \Delta_0 \, \sigma_3 \; 
  \right  \}  \psi (\vec {x}) \; = \; \varepsilon \;  \psi (\vec {x}) \; .
\end{equation}
Proceeding as in Appendix~\ref{Ap}, we get:
\begin{equation}
\label{4.8}
 \left ( \begin{array}{cc} 
+ \frac{\Delta_0}{v_F}   &   \hslash \partial_1 - i \frac{eB}{c} x_2 - i \hslash \partial_2 \\ 
  - \hslash \partial_1 + i \frac{eB}{c} x_2 - i \hslash \partial_2  &   - \frac{\Delta_0}{v_F}
\end{array}
\right )   \psi (\vec {x}) \; = \; \frac{\varepsilon}{v_F} \psi (\vec {x}) \; .
\end{equation}
Using   Eq.~(\ref{A.9}) we get:
\begin{equation}
\label{4.9}
 \begin{array}{c}
\left [ i p  - i \frac{eB}{c} x_2 - i \hslash \partial_2 \right ]   \phi_2(x_2) \; = \;  \frac{\varepsilon-\Delta_0}{v_F}     \phi_1(x_2)
 \\
\left [ - i p  + i \frac{eB}{c} x_2 - i \hslash \partial_2 \right ]   \phi_1(x_2) \; = \;  \frac{\varepsilon+\Delta_0}{v_F}     \phi_2(x_2) \; .
\end{array} 
\end{equation}  
From  Eq.~(\ref{4.9}) we easily obtain:
\begin{equation}
\label{4.10}
\begin{array}{c}
\left [-  \frac{\hslash^2}{2}  \partial_2^2  
 + \frac{1}{2}  \frac{e^2B^2}{c^2} (x_2 - \frac{p c}{eB})^2 \right ]  
 \phi_1(x_2) \; = \;  \frac{1}{2}
 \left [ \frac{\varepsilon^2-\Delta_0^2}{v_F^2} -  \frac{\hslash e B}{c} \right ]    \phi_1(x_2)
 \\
 \phi_2(x_2) \; =  \;  \frac{v_F}{\varepsilon+\Delta_0}   \left [ - i p  + i \frac{eB}{c} x_2 - i \hslash \partial_2 \right ]   \phi_1(x_2)  \; . 
\end{array} 
\end{equation}  
The solutions of  Eq.~(\ref{4.10}) are given in Eq.~(\ref{A.17}) with eigenvalues:
\begin{equation}
\label{4.11}
\varepsilon \; =\; \pm \,  \sqrt{ 2 \hslash \frac{eB}{c} (n \, + \, 1) v_F^2 \, + \, \Delta_0^2}  \; \; , \; \; n \, = \, 0, \, 1, \, ... 
\end{equation}
Obviously, we have also the zero modes with wavefunctions given by   Eq.~(\ref{A.18}) and eigenvalue: 
\begin{equation}
\label{4.12}
\varepsilon_0 \; = \; -  \;  \Delta_0 \; . 
\end{equation}
Therefore, we are lead to the following spectrum:
\begin{eqnarray}
\label{4.13}
\varepsilon_n^{(+)}  \; =  \;  + \,  \sqrt{ 2 \hslash \frac{eB}{c} \, v_F^2 \, n \,  + \, \Delta_0^2}  \; \; , \; \; n \, = \, 1, \, 2, \, ... 
\nonumber  \\
\varepsilon_n^{(-)}  \; = \;  -  \,  \sqrt{ 2 \hslash \frac{eB}{c} \, v_F^2  \, n \, + \, \Delta_0^2}  \; \; , \; \; n \, = \, 0, \, 1, \, ... 
\end{eqnarray}
where the  eigenfunctions are given by  Eq.~(\ref{A.21}).
\begin{figure}
\begin{center}
\includegraphics*[width=0.8\columnwidth]{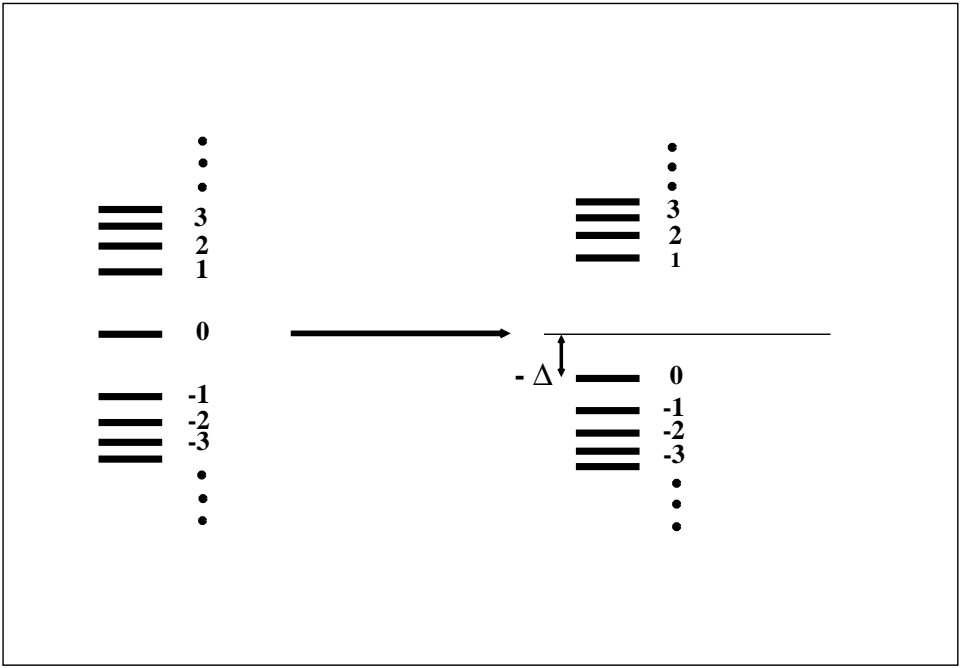}
\caption{\label{fig-1}
Schematic spectrum of Landau levels of surface states in topological insulators  in applied magnetic field (left).
 Landau levels with  dynamical generation of the gap $\Delta$ (right). }
\end{center}
\end{figure}
For convenience, in Fig.~\ref{fig-1} we compare the spectrum of the Landau levels with  $\Delta=0$ to the case
of non-zero gap. \\ 
Thus, we can write:
\begin{equation}
\label{4.14}
 \hat{\Psi}(\vec{x}, t) =  \int^{+\infty}_{-\infty} 
d p  \;  \left \{    \sum^{\infty}_{n=1}   e^{-i \frac{\varepsilon_{n} \; t}{\hslash} }  \; 
 \psi^{(+)}_{n,p}(\vec{x}) \; \hat{b}_{n,p}   \;   +  \; \sum^{\infty}_{n=0} e^{+ i \frac{\varepsilon_{n} \; t}{\hslash} }  \;
 \psi^{(-)}_{n,p}(\vec{x}) \; \hat{d}^{\dagger}_{n,p}  \; 
  \right \}  \; ,
\end{equation}
\begin{equation}
\label{4.15}
\hat{\Psi}^{\dagger}(\vec{x}, t)  =  \int^{+\infty}_{-\infty}    dp \; 
\left \{   
  \sum^{\infty}_{n=1} e^{+i \frac{\varepsilon_{n} \; t}{\hslash} }  \; 
 \psi^{(+)\dagger}_{n,p}(\vec{x}) \; \hat{b}^{\dagger}_{n,p}   \;  + \;  \sum^{\infty}_{n=0}  e^{- i \frac{\varepsilon_{n} \; t}{\hslash} }  \;
 \psi^{(-)\dagger}_{n,p}(\vec{x}) \; \hat{d}_{n,p}  \;  
  \right \}  \; ,
\end{equation}
where the creation and annihilation operators operators satisfy the standard  anticommutation relations, and
\begin{equation}
\label{4.16}
\varepsilon_n  \; =  \;   \sqrt{ 2 \hslash \frac{eB}{c} \, v_F^2 \, n \,  + \, \Delta_0^2}  \; \;  . 
\end{equation}
Whereupon, we obtain  the Hamiltonian operator:
\begin{equation}
\label{4.17}
\hat{H}  \;  =  \;   \int^{+\infty}_{-\infty}  d p \;    \left \{  \;   \sum^{\infty}_{n=1}     
 \varepsilon_{n} \; \hat{b}^{\dagger}_{n,p}  \,  \hat{b}_{n,p}  \; +  \;   \sum^{\infty}_{n=0}      
 \varepsilon_{n} \;  \hat{d}^{\dagger}_{n,p} \,  \hat{d}_{n,p}  \right  \}   +  E_0(B)  \; , 
\end{equation}
with:
\begin{equation}
\label{4.18}
E_0(B) \;  = \; - \; V \;  \frac {eB} {2 \pi \hslash c}  \; \sum^{\infty}_{n=0}  \,   \varepsilon_{n} \; .
\end{equation}
The renormalized Hamiltonian operator is:
\begin{equation}
\label{4.19}
\hat{H}_R  \;  =  \;   \int^{+\infty}_{-\infty}  d p \;    \left \{  \;   \sum^{\infty}_{n=1}     
 \varepsilon_{n} \; \hat{b}^{\dagger}_{n,p}  \,  \hat{b}_{n,p}  \; +  \;   \sum^{\infty}_{n=0}      
 \varepsilon_{n} \;  \hat{d}^{\dagger}_{n,p} \,  \hat{d}_{n,p}  \right  \}   +  \Delta E_0(B)  \; , 
\end{equation}
where now:
\begin{equation}
\label{4.20}
\Delta E_0(B) \;  = \; - \; V \;  \frac {eB} {2 \pi \hslash c}  \; \sum^{\infty}_{n=0}  \,    
 \sqrt{ 2 \hslash \frac{eB}{c} \, v_F^2 \, n \,  + \, \Delta_0^2}  \; - \; E_0 \;  .
\end{equation}
%
\begin{figure}
\begin{center}
\includegraphics*[width=0.8\columnwidth]{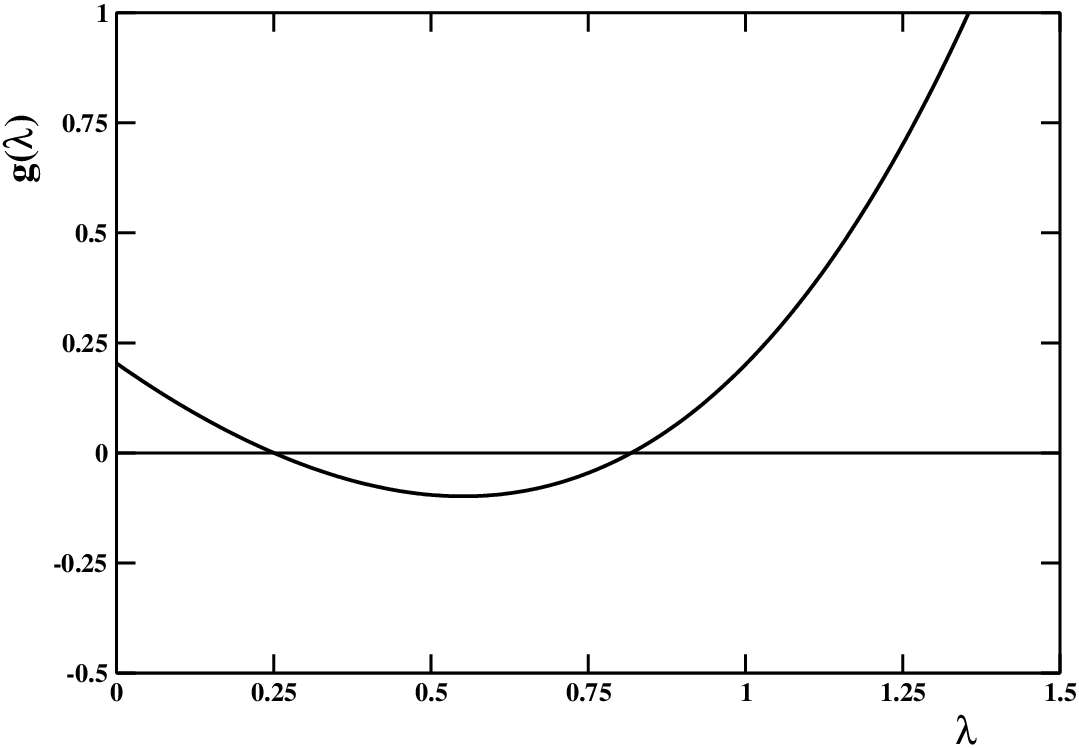}
\caption{\label{fig-2}
Plot of the function  $g(\lambda)$, Eq.~(\ref{4.27}),  versus $\lambda$.}
\end{center}
\end{figure}
%
To evaluate the (divergent) sum in   Eq.~(\ref{4.20}) we employ the same trick as before. After some algebra it is easy to show that:
\begin{equation}
\label{4.21}
 \sum^{\infty}_{n=0}  \,   \sqrt{ 2 \hslash \frac{eB}{c} \, v_F^2 \, n \,  + \, \Delta_0^2}  \; =  \;  
 + \, \Delta_0 \; - \;   \int_0^{\infty}  \frac {ds} {\sqrt{\pi s}}  \frac{d}{ds} 
 \left [    \frac{e^{-\Delta_0^2 s}}{ e^{  + 2 \hslash v_F^2  \frac{eB}{c}  s} - 1} \;   \right ] \; .
\end{equation}
Since
\begin{equation}
\label{4.22}
 \frac{e^{-\Delta_0^2 s}}{ e^{  + 2 \hslash v_F^2  \frac{eB}{c}  s} - 1}  
 \;  \stackrel{ s  \to  0}{\sim}  \; 
 + \;  \frac {1} {   2 \hslash v_F^2  \frac{eB}{c}  s    }  \;  \;  \; ,
\end{equation}
we  readily   obtain:
\begin{equation}
\label{4.23}
\Delta E_0(B) \;  = \;  V \;  \frac {eB} {2 \pi \hslash c}  \;   \left \{
 - \, \Delta_0 \; + \;   \int_0^{\infty}  \frac {ds} {\sqrt{\pi s}}  \frac{d}{ds} 
 \left [    \frac{e^{-\Delta_0^2 s}}{ e^{  + 2 \hslash v_F^2  \frac{eB}{c}  s} - 1}  \;   
 - \;  \frac {1} {   2 \hslash v_F^2  \frac{eB}{c}  s    }  \;  \right ] \; 
 \right \}  \; .
\end{equation}
Therefore the vacuum energy per particle is:
\begin{equation}
\label{4.24}
\varepsilon_P(B, \Delta_0) \;  = \; 
 - \, \Delta_0 \; + \;   \int_0^{\infty}  \frac {ds} {\sqrt{\pi s}}  \frac{d}{ds} 
 \left [    \frac{e^{-\Delta_0^2 s}}{ e^{  + 2 \hslash v_F^2  \frac{eB}{c}  s} - 1}  \;   
 - \;  \frac {1} {   2 \hslash v_F^2  \frac{eB}{c}  s    }  \;  \right ]  \;   \; .
\end{equation}
Introducing the dimensionless variable:
\begin{equation}
\label{4.25}
\lambda  \;  = \;  
 \frac {\Delta_0} {   \sqrt{2 \hslash v_F^2  \frac{eB}{c}  } }  \;    \; ,
\end{equation}
Eq.~(\ref{4.24}) can be rewritten as:
\begin{equation}
\label{4.26}
\varepsilon_P(B, \Delta_0) \;  = \; 
 \sqrt{ 2 \hslash v_F^2  \frac{eB}{c}  }  \; \;  g(\lambda)   \;  \; ,
\end{equation}
where:
\begin{equation}
\label{4.27}
 g(\lambda)  \;  = \; 
 - \, \lambda \; + \;   \int_0^{\infty}  \frac {dx} {\sqrt{\pi x}}  \frac{d}{dx} 
 \left [    \frac{e^{-\lambda^2 x}}{ e^{ x} - 1}  \;    - \;  \frac {1} { x  }  \;  \right ]  \;   \; .
\end{equation}
It is straightforward  to see that the function $g(\lambda)$ is finite. Indeed, in Fig.~\ref{fig-2} we display $g(\lambda)$ versus $\lambda$.
Eqs.~(\ref{4.26}) and   (\ref{4.27}) imply that the vacuum energy per particle   depends not only
 on the magnetic field but also on the gap $\Delta_0$, which until now has been  a free positive
parameter.  It is worthwhile to stress that  for  $\Delta_0=0$    Eq.~(\ref{4.26}) agrees with  Eq.~(\ref{3.15}) since $g(0) \simeq 0.208$. 
Looking at   Fig.~\ref{fig-2} we note that the vacuum energy per particle decreases for small values of the gap $\Delta_0$, reaches
a minimum, and then increases for large enough values of the gap. In fact, we find the minimum at 
 $\lambda = \overline \lambda \simeq 0.56$  with  $g(\overline \lambda) \simeq  - 0.093$. 
To understand this peculiar  behavior of the vacuum energy, we observe that the opening of the gap $\Delta_0$ affects the spectrum
of the Landau levels in two way. On the  one hand  the gap decreases the energy of the zero modes (see Fig.~\ref{fig-1}). On the other hand
the energy of higher Landau levels  tend to increase with the gap. It is the balancing of these two effects   that results in the
vacuum energy minimum.
Thus, we see that it is energetically favorable to induce a dynamical gap by a rearrangement of the Dirac sea  triggered
by quantum fluctuations. This remarkable phenomenon is quite analogous to the  Peierls~\cite{Peierls:1955} quantum phase transition
in quasi one-dimensional conductors. \\
In summary, we found that the low-lying surface states in three-dimensional topological insulators immersed in an external
magnetic field give rise to gapped Landau levels:
\begin{eqnarray}
\label{4.28}
\varepsilon_n^{(+)}  \; =  \;  E_D(B)  \; + \,  \sqrt{ 2 \hslash \frac{eB}{c} \, v_F^2 \, n \,  + \, \Delta_0^2}  \; \; , \; \; n \, = \, 1, \, 2, \, ... 
\nonumber  \\
\varepsilon_n^{(-)}  \; = \; E_D(B) \;   -  \,  \sqrt{ 2 \hslash \frac{eB}{c} \, v_F^2  \, n \, + \, \Delta_0^2}  \; \; , \; \; n \, = \, 0, \, 1, \, ... 
\end{eqnarray}
with:
\begin{equation}
\label{4.29}
 \Delta_0 \; = \overline \lambda \;  \sqrt{ 2 \hslash  v_F^2  \frac{ eB}{c} } \;  \simeq \; 
 0.56 \;  \sqrt{ 2 \hslash  v_F^2  \frac{ eB}{c} }  \;  ,
\end{equation}
and
\begin{equation}
\label{4.30}
 E_D(B) \; \equiv \;  E_D \; + \; \varepsilon_P(B, \Delta_0) \;  \simeq \; E_D \;  - \, 0.093 \; 
 \sqrt{ 2 \hslash v_F^2  \frac{eB}{c}  }  \; \; .
\end{equation}
The results obtained in this Section rely on our effective Hamiltonian which  should describe the low-lying
excitations of the surface states in topological insulators. One must still face the task of comparing
the results obtained within a highly-idealized model with experimental investigations of three-dimensional
topological insulators.
%
\section{Landau Levels: Comparison with  Experimental Data}
\label{S5}
In the preceding Section we argued that it is energetically favored by quantum fluctuations to develop a mass gap for
the low-lying surface excitations in three-dimensional topological insulator in an external magnetic field.
The aim of this Section is to contrast our theoretical expectations to experimental data. \\
Quite recently, several experimental studies confirmed that the peculiar relativistic nature of surface states in topological insulators
in magnetic fields leads to  Landau levels which varies linearly with $\sqrt{n B}$.
The authors of Ref.~\cite{Cheng:2010a,Cheng:2010b}    reported direct observation of Landau quantization of the topological surface states
in Bi$_2$Se$_3$ in magnetic fields by using scanning tunneling microscopy and spectroscopy.
Relativistic Landau levels are also observed in the tunneling spectra in a magnetic field~\cite{Hanaguri:2010} 
by using scanning tunneling spectroscopy on the surface of the topological insulator.    
Moreover,  microwave spectroscopy has been applied to study cyclotron resonance due to intraband transitions between Landau levels
in  Bi$_2$Se$_3$~\cite{Wolos:2012}.  \\
The hallmark of Dirac fermions for  low-lying surface excitations  in three-dimensional topological insulator
is   the presence of a field independent Landau level  at the Dirac point (zero mode) and the already alluded scaling of higher Landau
levels with   $\sqrt{n B}$. In fact, Fig.~4 in Ref.~\cite{Cheng:2010a} displays the energies of the Landau levels versus 
$\sqrt{n B}$ for magnetic fields ranging from  8 T  to 11 T~\footnote{Even though we are using cgs units, it is widespread practice  to 
measure the strength of the magnetic field in Tesla. We recall that 1 T = $10^4$ G.}. Remarkably, for higher values of n one observes Landau
level energies which vary linearly with  $\sqrt{n B}$. Moreover,  from  Fig.~2 in Ref.~\cite{Cheng:2010a} one infers that the
lowest Landau level, assumed to be the zero mode  $n=0$, is pinned at the Dirac point:
\begin{equation}
\label{5.1}
 E_D  \; \simeq  \;  -200 \; mev \; \; ,  
\end{equation}
consistently with the value inferred from the minimum of the differential tunneling conductance in absence of magnetic field.
However, for low $n$, namely for the Landau levels with energies near the Dirac point, it seems that the Landau levels deviate
from the linear behavior.  We believe that this effect can be accounted for if we admit the presence of a dynamical mass gap
according to our previous  discussion in Sect.~\ref{S4}. To check quantitatively this, we need the energies of the Landau levels for
different magnetic field strength. 
Fortunately,  Fig.~2 in Ref.~\cite{Cheng:2010b} reports the Landau levels for different magnetic fields ranging from 6 T up to 11 T.
We used  the experimental data extracted from  Fig.~2 of Ref.~\cite{Cheng:2010b} to perform a quantitative comparison with
the theoretical results discussed in the previous Section. To this end, we note that in  Refs.~\cite{Cheng:2010a,Cheng:2010b}
the energy of the Landau levels are determined by fitting the differential conductance with multiple gaussians. The typical
peak width of Landau levels turns out to be of about 8 mev. Therefore, to be conservative, for our analysis we assumed for
the Landau level energies a statistical uncertainty of about   2 - 3 mev. \\
Preliminarily, to check our procedure, we fitted the data to the expected spectrum of Landau levels for massless Dirac fermions, Eq.~(\ref{1.1}).
Since the average Fermi velocity in   Bi$_2$Se$_3$  lies in the range $v_F = 3.0 - 7.0  \times 10^{7}$  cm/s~\cite{Kim:2012},
we write $v_F = v_5  \times 5.0 \; 10^7$ cm/s. Accordingly,  for the spectrum lying above the Dirac point, Eq.~(\ref{1.1}) can be rewritten as:
\begin{equation}
\label{5.2}
\varepsilon^{(+)}_n \; =  \; E_D \; +  \;   18.34 \; mev \; v_5 \; \sqrt{ n \; B(T)}  \;  \;  , \;  \;  n \;  =  \;  0 \, , \,  1 \, , \,  2  \, , \;  . \, . \, ,
\end{equation}
where B(T) stands for magnetic field strength in Tesla.  \\
To implement  the fits  we used  the program Minuit~\cite{James:1975} which is conceived as a tool to find the minimum value of a multi-parameter   
 function and analyze the shape of the function around the minimum. The principal application is to compute the best-fit parameter values and
 uncertainties by minimizing the total chi-square $\chi^2$. \\
\begin{figure}
\begin{center}
\includegraphics*[width=0.6\columnwidth]{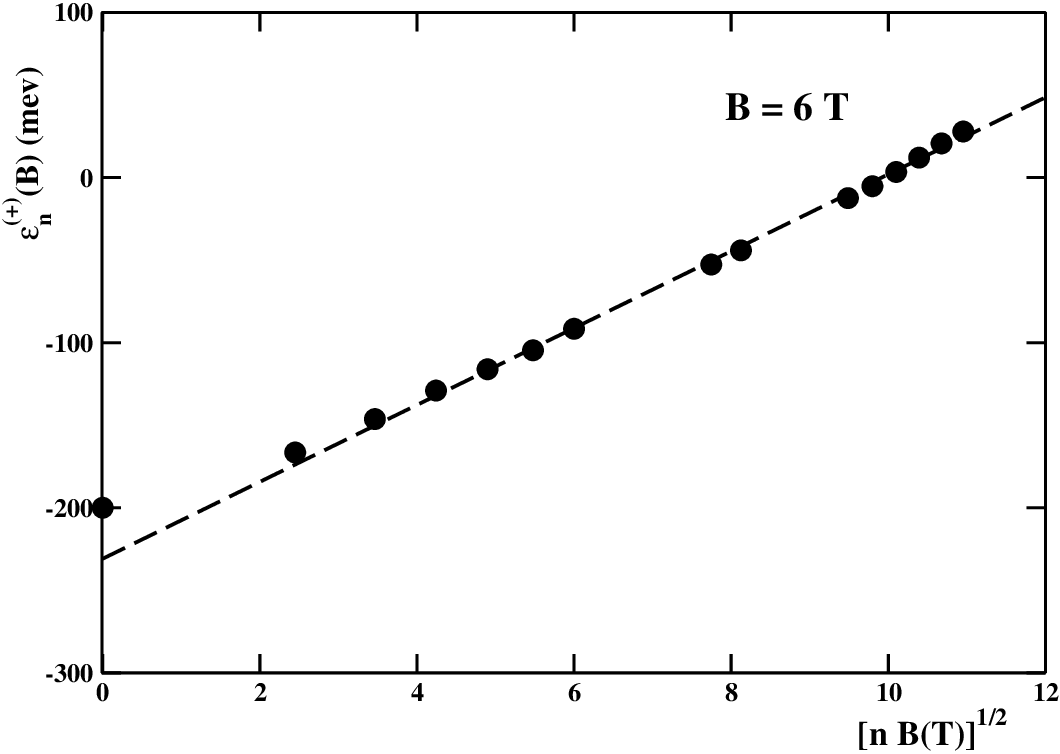}
\\
\vspace{.01\columnwidth}
\includegraphics*[width=0.6\columnwidth]{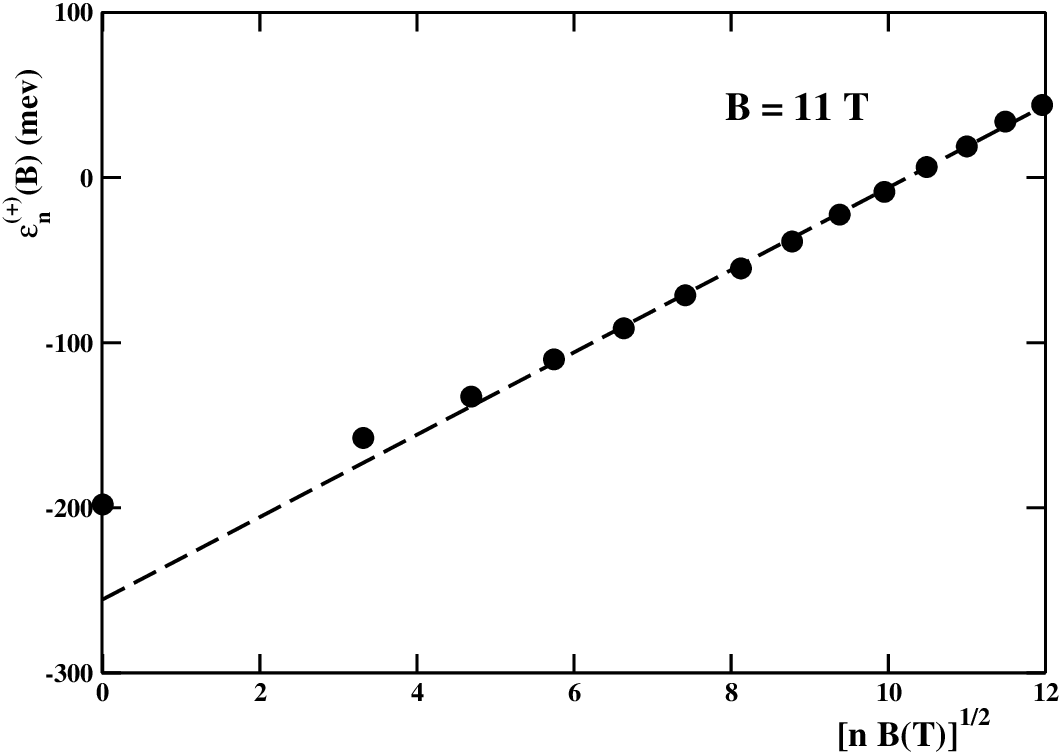}
\caption{\label{fig-3} Landau level energies $\varepsilon_n^{(+)}(B)$ versus $\sqrt{nB}$ for B = 7 T (Top) and  B = 11 T (Bottom).
Data have been  borrowed from  Fig.2 of Ref.~\cite{Cheng:2010b}. The assumed statistical errors are smaller than the symbol
size. Dashed lines are the fits of $\varepsilon_n^{(+)}(B)$ to Eq.~(\ref{5.2}).}
\end{center}
\end{figure}
%
%
We fitted  the experimental data to Eq.~(\ref{5.2})  for each values of the magnetic field leaving $E_D$ and $v_5$ as free fitting parameters.
Moreover, to get a sensible fit to the data we excluded the  Landau levels with $n=0,1$.  In Table~\ref{Table-1}
we summarize the results of the fitting procedure. Note that in the last column of  Table~\ref{Table-1} we indicate the reduced
chi-square, namely the total chi-square $\chi^2$ divided by the number of degree of freedom. As  rule of thumb,  a sensible fit
results in $\chi_r^2 \sim 1$. \\
%
\begin{table}[t]
\setlength{\tabcolsep}{0.9pc}
\centering
\caption[]{Summary of the values of  fitting parameters in  Eq.~(\ref{5.2})
reported, respectively, in the second and third columns.
In the  fourth column we give the reduced chi-square.}
\begin{tabular}{cllll}
\hline
\hline
B(T)  &  $E_D$ (mev)  &  $v_5$  & $\chi^2_r$    \\
\hline
  6   & -231 $\pm$  5 & 1.27 $\pm$ 0.03  & 1.20 \\
\hline
7  & -237 $\pm$  5 & 1.29 $\pm$ 0.03  & 1.00 \\
\hline
8 & -249 $\pm$  5 & 1.32 $\pm$ 0.03  & 1.05 \\
\hline
9  & -246 $\pm$  5 & 1.30 $\pm$ 0.03  & 0.77 \\
\hline
10 & -251 $\pm$  5 & 1.32 $\pm$ 0.03  & 0.95 \\
\hline
11 & -256 $\pm$  6 & 1.36 $\pm$ 0.03  & 0.99 \\
\hline
\hline
\end{tabular}
\label{Table-1}
\end{table}
%
In Fig.~\ref{fig-3} we display the experimental data together  with the fitting lines for two representative values of the magnetic field.
As it turns out, the lowest Landau level has an energy quite close to the Dirac point $E_D \simeq$  - 200 mev independently on the
magnetic field strengths. However, the lowest Landau levels $n=0,1$ seems to deviate appreciably from the linear behavior  $\sim \sqrt{nB}$
which fits nicely the  $n \ge 2$ Landau levels.  Even more, if we extrapolate $\varepsilon^{(+)}_n$   according to Eq.~(\ref{5.2}) we obtain 
the energies of the Dirac point,  showed in Table~\ref{Table-1},  which, not only  lie  very far from the value at zero magnetic field,  Eq.~(\ref{5.1}),
but seem to decrease with the magnetic field. On the other hand, the fitted values of the Fermi velocity $v_F$ does not show a sizable dependence
on the magnetic field strengths. In fact, in   Fig.~\ref{fig-4} we report the fitted values of the Fermi velocity versus the magnetic field strength. 
We find for the average Fermi velocity:
\begin{equation}
\label{5.3}
 v_F  \;  =   \; 6.55 \; \pm \; 0.12 \; 10^7 \; cm/s \; \; \; , \; \; \; \chi^2_r \;  =   \; 1.07  \; \; .  
\end{equation}
It is reassuring to see that $v_F$ in Eq.~(\ref{5.3})  lies in allowed  range for the average Fermi velocity in   Bi$_2$Se$_3$~\cite{Kim:2012}.
\\
%
\begin{figure}
\begin{center}
\includegraphics*[width=0.8\columnwidth]{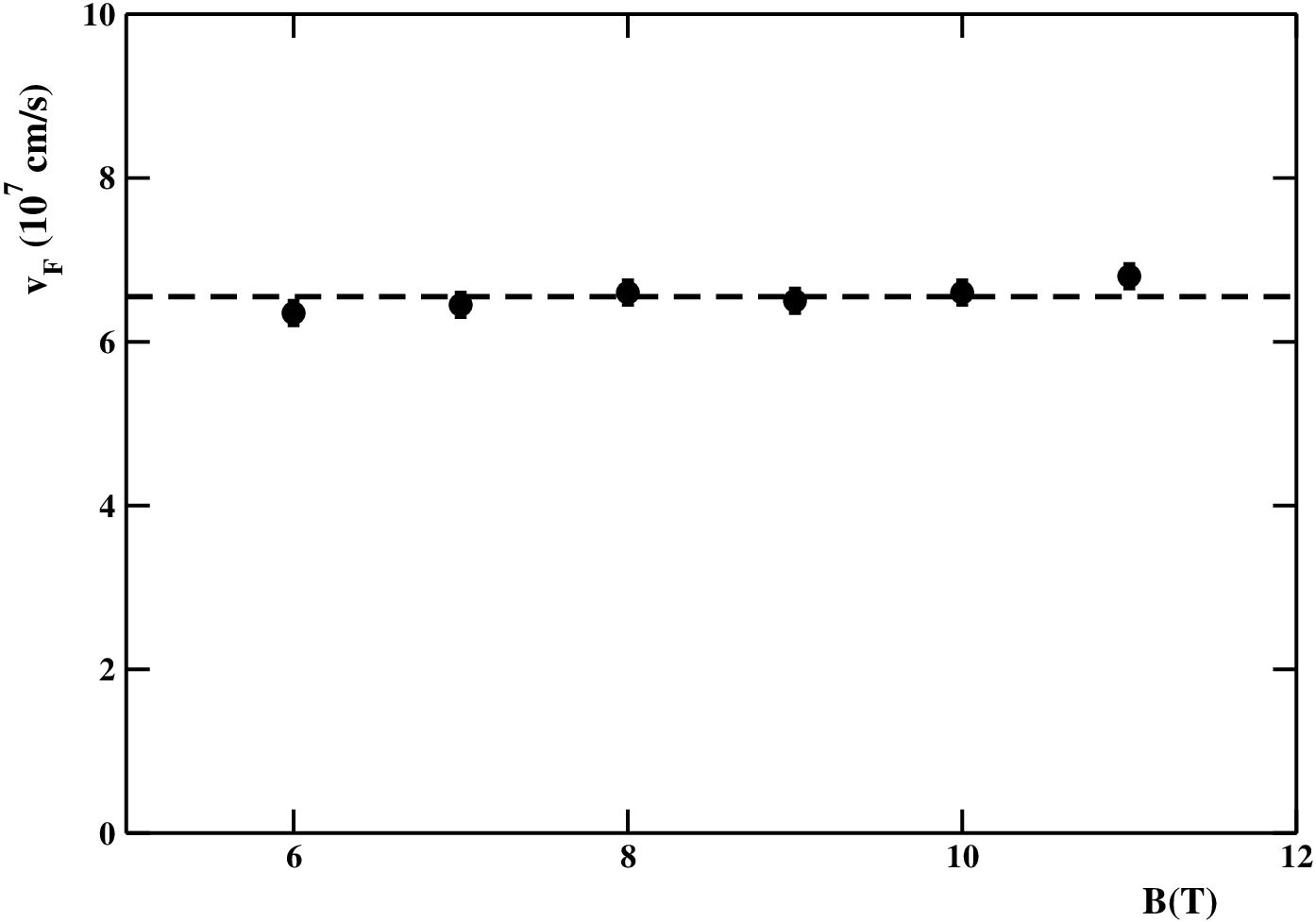}
\caption{\label{fig-4}
Fermi velocity in $10^7 \; cm/s$  versus the magnetic field B(T). 
The dashed line is the  average velocity, Eq.(\ref{5.3}), obtained by fitting with a constant the values in Table~\ref{Table-1}.}
\end{center}
\end{figure}
%
%
The puzzling anomalies encountered within the standard interpretation of the measured Landau level energies point to
an alternative interpretation of the experimental data. We shall, now,  show  that the measurements can be accounted for
in a consistent manner within the theoretical scenario discussed in Sect.~\ref{S4}. In this case, the spectrum of the Landau levels
is given by Eqs.~(\ref{4.28}), (\ref{4.29}), and (\ref{4.30}).
To this end, we need the Landau levels which are above the Dirac point. According to  Eq.~(\ref{4.28}) we have:
\begin{equation}
\label{5.4}
\varepsilon_n^{(+)}  \; =  \;  E_D(B)  \; + \,  \sqrt{ 2 \hslash \frac{eB}{c} \, v_F^2 \, n \,  + \, \Delta_0^2}  \; \; , \; \; n \, = \, 1, \, 2, \, ...  \; ,
\end{equation}
with:
\begin{equation}
\label{5.5}
 \Delta_0 \; = \delta \;  \sqrt{ 2 \hslash  v_F^2  \frac{ eB}{c} } \; \; , \; \; \delta \; \simeq \; 
 0.56 \;  \;  ,
\end{equation}
and
\begin{equation}
\label{5.6}
 E_D(B) \; =    \; E_D \;  - \;  \alpha \;  \sqrt{ 2 \hslash v_F^2  \frac{eB}{c}  }  \; \; , \; \; \alpha \; \simeq \; 0.093  \; .
\end{equation}
For convenience we rewrite  Eq.~(\ref{5.4})  as:
\begin{equation}
\label{5.7}
\varepsilon^{(+)}_n \; =  \; E_D(B)  \;  +  \;   18.34 \; mev \; v_5 \; \sqrt{ B(T) \; (n \; + \; \delta^2)}  \;  \;  , \;  \;  n \;  =  \;  1 \, , \,  2  \, , \;  . \, . \, .  \; .
\end{equation}
%
\begin{table}[thb]
\setlength{\tabcolsep}{0.9pc}
\centering
\caption[]{Summary of the values of the fitting parameters in  Eq.~(\ref{5.7})
reported, respectively, in the second, third  and fourth columns.
In the  last column we give the reduced chi-square.}
\begin{tabular}{clllll}
\hline
\hline
B(T)  &  $E_D(B)$ (mev)  &  $v_5$  &  $\delta$  & $\chi^2_r$    \\
\hline
  6   & -262.0 $\pm$  4.2 & 1.39 $\pm$ 0.03 & 0.50 $\pm$ 0.36 & 1.33 \\
\hline
7  & -271.1 $\pm$  4.4 &  1.42 $\pm$ 0.02  & 0.50 $\pm$ 0.36 & 1.21 \\
\hline
8 & -283.5 $\pm$  4.4 & 1.44 $\pm$ 0.03  & 0.50 $\pm$ 0.42 & 0.92 \\
\hline
9  & -288.9 $\pm$  4.5 & 1.45 $\pm$ 0.02  & 0.50 $\pm$ 0.35 & 0.67 \\
\hline
10 & -298.1 $\pm$  4.6 & 1.48 $\pm$ 0.02 & 0.50 $\pm$ 0.35  & 0.76 \\
\hline
11 & -299.3 $\pm$  4.9 & 1.49 $\pm$ 0.08 & 0.55 $\pm$ 0.13  & 0.42 \\
\hline
\hline
\end{tabular}
\label{Table-2}
\end{table}
\begin{figure}
\begin{center}
\includegraphics*[width=0.6\columnwidth]{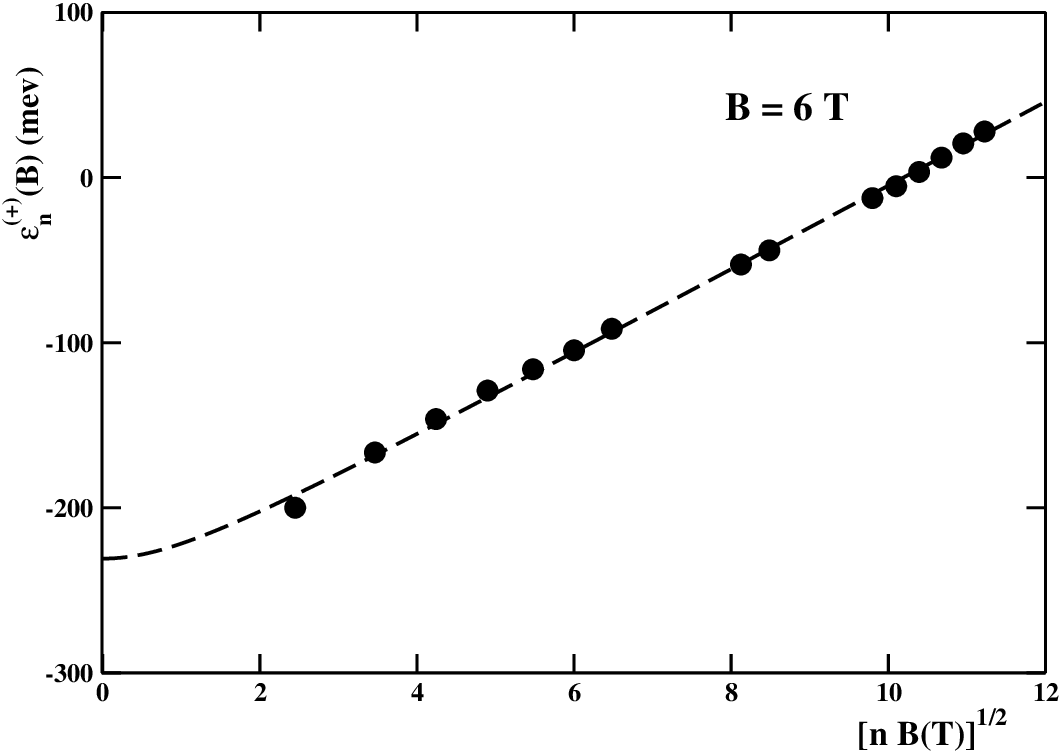}
\\
\vspace{0.003\columnwidth}
\includegraphics*[width=0.6\columnwidth]{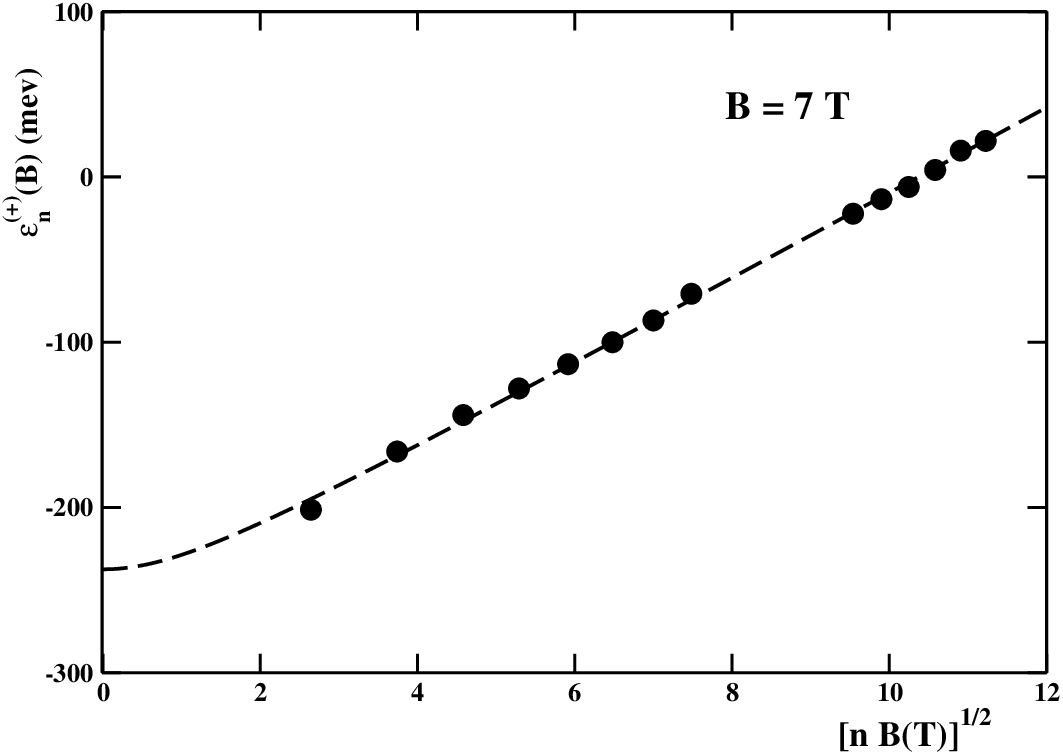}
\\
\vspace{0.003\columnwidth}
\includegraphics*[width=0.6\columnwidth]{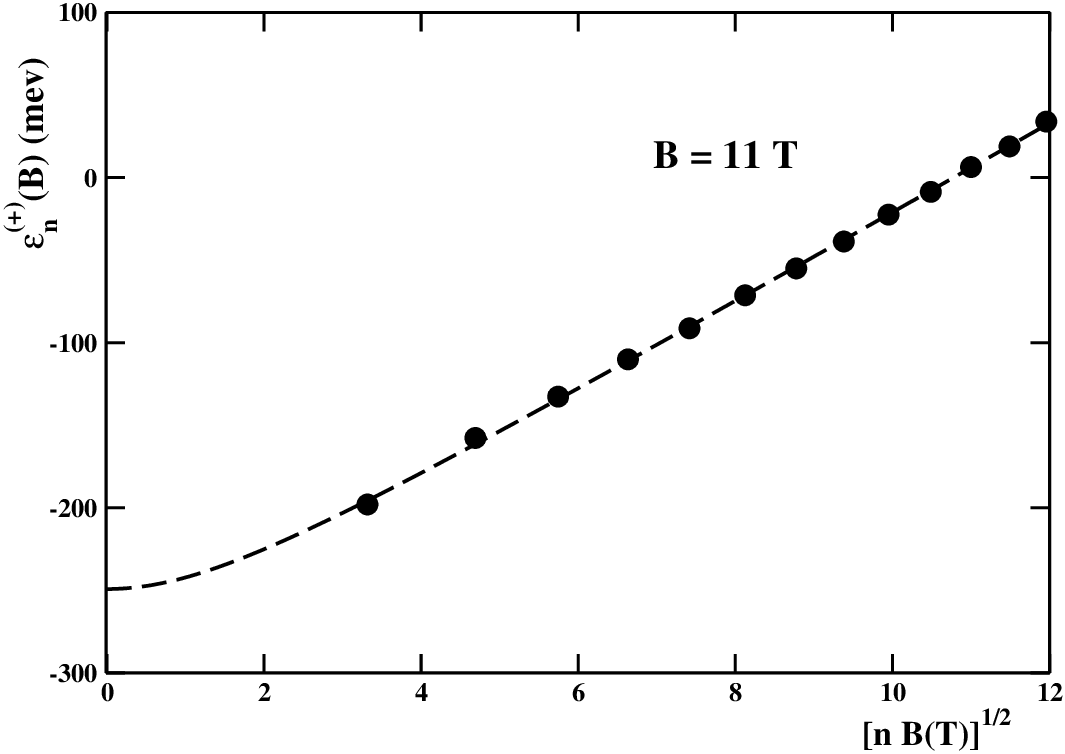}
\caption{\label{fig-5}
Landau level energies $\varepsilon^{(+)}_n $ versus $\sqrt{nB}$ for B = 6 T (Top),  B = 7 T (Middle), and  B = 11 T (Bottom). 
Data have been extracted form Fig.2 of Ref.~\cite{Cheng:2010b}. The assumed statistical errors are smaller than the symbol
size. Dashed lines are the fits of $\varepsilon_n^{(+)}(B)$ to Eq.~(\ref{5.7}).}
\end{center}
\end{figure}
We have fitted the available data to   Eq.~(\ref{5.7}) leaving $E_D(B)$, $v_5$, and $\delta$ as free parameters. As a result, the fitting
procedure  gives the best fit values of the parameter for each given value of the magnetic field. For convenience, the results of our fits
are summarize in Table~\ref{Table-2}, while  in Fig.~\ref{fig-5} we display the experimental data together 
 with the fitting curves for three representative values of the magnetic field.
\\
A few comments are in order. First, we note that the Dirac energy $E_D(B)$ do depend on the magnetic field as in the previous fits. However,
presently we expect a shift of the Dirac point according to  Eq.~(\ref{5.6}). In fact, as  we  will discuss later the shift of the Dirac point due
to the magnetic field seems to be consistent with theoretical expectations.  From Table~\ref{Table-2} we see that the Fermi velocity is
independent on the magnetic field strengths. In fact, we find for the average Fermi velocity:
\begin{equation}
\label{5.8}
 v_F  \;  =   \; 7.2 \; \pm \; 0.1  \; 10^7 \; cm/s \; \; \; , \; \; \; \chi^2_r \;  =   \; 1.67  \; \; .
\end{equation}
This value of the Fermi velocity is slightly higher, but in reasonable agreement with our previous determination  Eq.~(\ref{5.3}).  \\
As concern the spectrum of the Landau levels, we stress that the levels lying above the Dirac point have $n \ge 1$, according to
Eq.~(\ref{5.7}). In fact, the zero mode $n=0$ has been pushed below $E_D(B)$ due to the dynamical generation of the mass gap.
Therefore,  Fig.~\ref{fig-5} shows that our theoretical curves are able to track all the experimental data. Note that the lowest
Landau level $n=1$ has an energy which seems to be almost independent on the magnetic field. This is due to the 
compensation of two different effects. Indeed, the increase of the energy due to the magnetic field for $n \neq 0$  is contrasted by
the negative shift of the Dirac point. For $n=1$ these two effects almost perfectly compensate each other. 
From  Fig.~\ref{fig-5} and Table~\ref{Table-2} we may conclude that the data are in reasonable agreement with 
Eq.~(\ref{5.7}) for all the available values of the magnetic field. Note that in Fig.~\ref{fig-5}, to magnify the non linearity
in $\sqrt{nB}$ due to the gap, the fitting curves have been draw starting from $nB=0$.  
Concerning the parameter $\delta$, looking at Table~\ref{Table-2} we see that this parameter does not depend on the magnetic field.
We find for the average value: 
\begin{equation}
\label{5.9} 
 \delta  \;  =   \; 0.51 \; \pm \; 0.18    \; \; \; , \; \; \; \chi^2_r \;  =   \; 0.01  \; \; ,
\end{equation}
which is in remarkable agreement with our theoretical expectations. \\
\begin{figure}
\begin{center}
\includegraphics*[width=0.8\columnwidth]{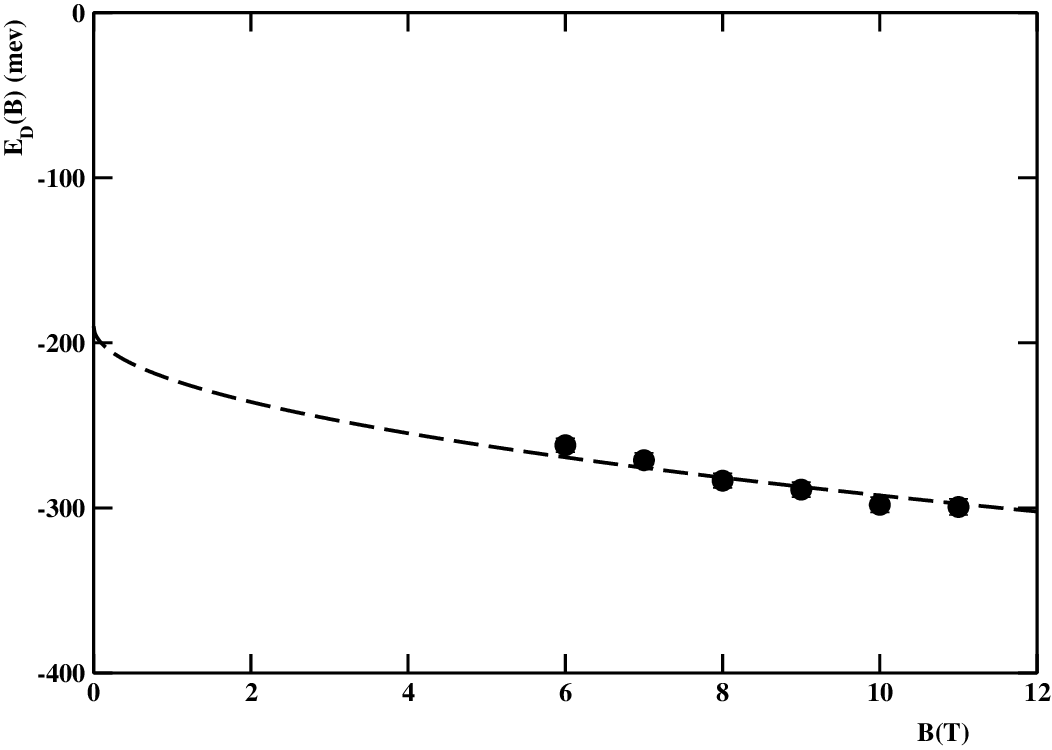}
\caption{\label{fig-6}
Dirac point energies $E_{D}$  in meV versus the magnetic field B in Tesla. 
Dashed line is  the fit of $E_{D}(B)$ to Eq.(\ref{5.6}).}
\end{center}
\end{figure}
%
%
Finally, we must check if the dependence of the Dirac energy on the magnetic field is in agreement with  Eq.~(\ref{5.6}).  
In  Fig.~\ref{fig-6} we report the fitted values of the Dirac energy in Table~\ref{Table-2} as a function of the magnetic field.
We have fitted these values according to Eq.~(\ref{5.6}) assuming for the Fermi velocity the average value  Eq.~(\ref{5.8}),
and leaving $E_D$ and $\alpha$ as free parameters. The fit returns for these two parameters the following values:
\begin{equation}
\label{5.10}
E_D   \;  =   \; - \;  190  \; \pm \;  14    \; mev  \; \; , \; \; \; \alpha \;  =   \; 1.22  \; \pm \;  0.04   \; \;  , \; \; \; \chi^2_r \;  =   \; 1.53 \; \; .
\end{equation}
Indeed, we find that the shift of the Dirac point with the magnetic field is in satisfying agreement with Eq.~(\ref{5.6}) 
(see dashed line in Fig.~\ref{fig-6}).
Moreover, the parameter $E_D$, which is the energy of the Dirac point in absence of the magnetic field, is in good agreement with
the experimental value  Eq.~(\ref{5.1}). However, we must stress that the parameter $\alpha$ is well above our theoretical
estimate,  Eq.~(\ref{5.6}).  We recall that this parameter quantify the vacuum energy per particle. In our opinion, since the calculations
in previous Section have been performed within a highly idealized model, such a discrepancy does not spoil the overall coherent
agreement of our theoretical picture with the experimental observations.
\section{Quantum Hall Effect and Chiral Edge States}
\label{S6}
\begin{figure}
\begin{center}
\includegraphics*[width=0.8\columnwidth]{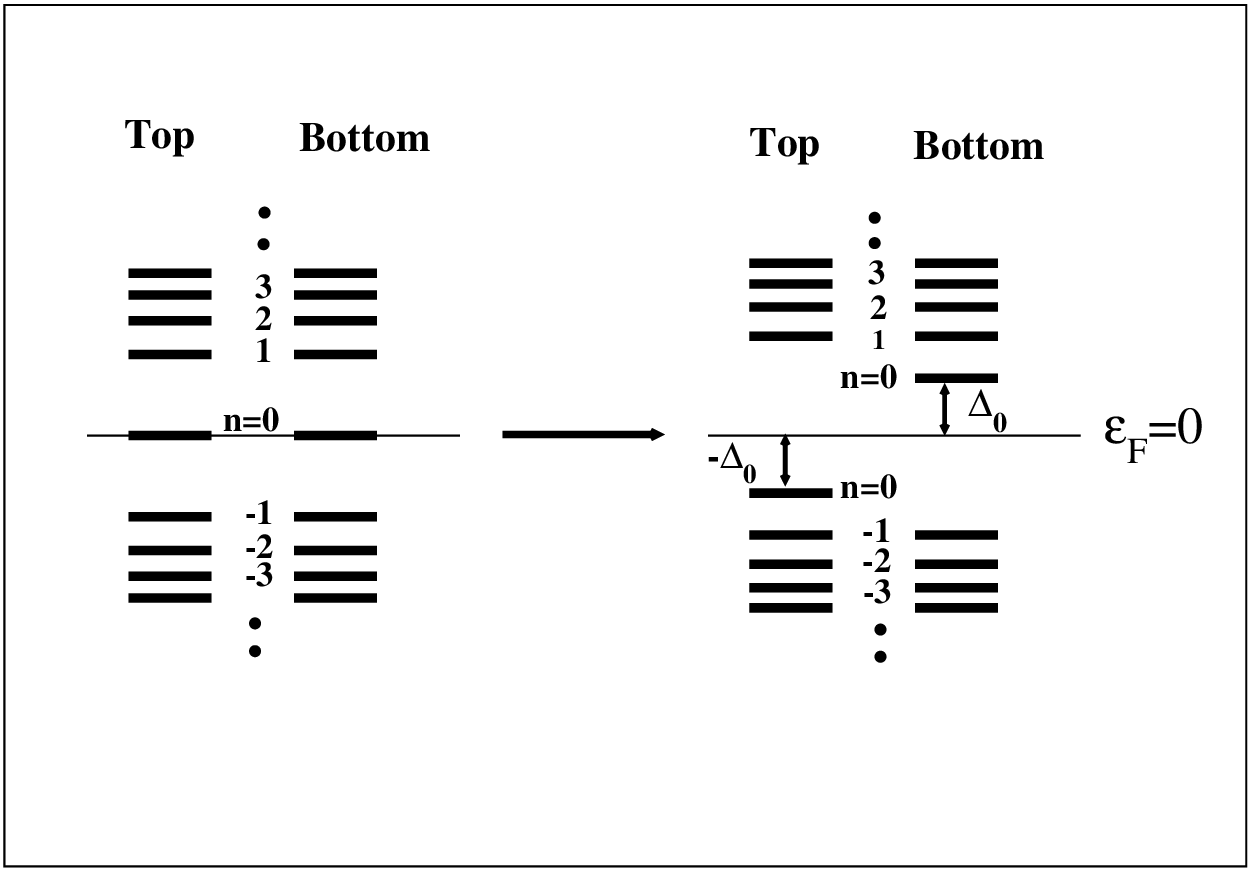}
\caption{\label{fig-7}
Schematic spectrum of surface state Landau levels in applied magnetic field  (left). Landau levels with  
dynamical generation of a gap $\Delta_0$ (right).  The Fermi level is at Dirac neutral point $\varepsilon_F =  0$.}
\end{center}
\end{figure}

\begin{figure}
\begin{center}
\includegraphics*[width=0.8\columnwidth]{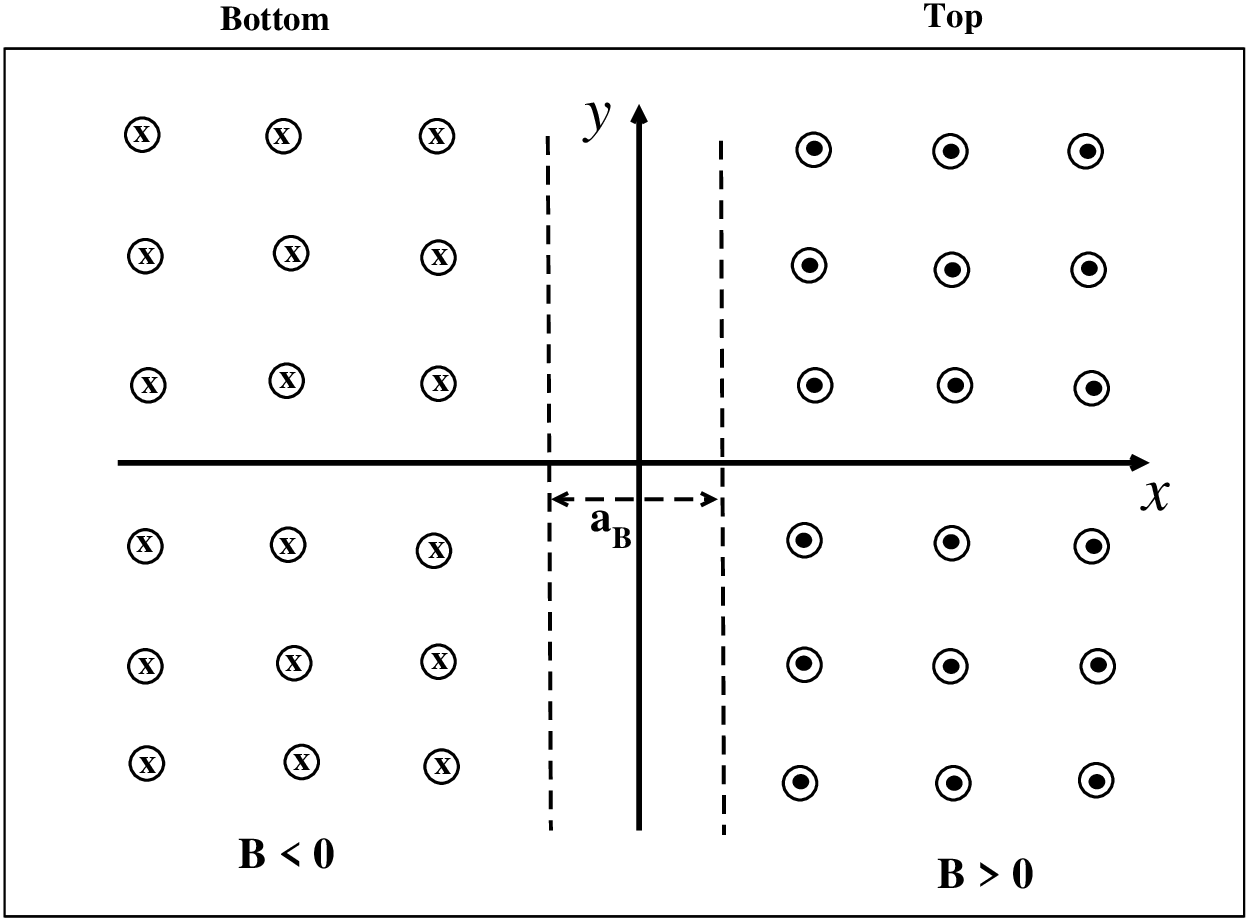}
\caption{\label{fig-8}
Schematic picture of the magnetic kink relevant for the edge states. Dots and crosses indicate the direction of magnetic 
field leaving and entering the surface, respectively. The width of the strip (dashed lines) where the magnetic field
change sign is set by the magnetic length  $a_B  =  \sqrt{\frac{\hslash c}{e B_0} } $. }
\end{center}
\end{figure}

The Hall effect indicates the voltage drop across a conductor transverse to the direction of the applied electrical current
in the presence of a perpendicular magnetic field. In the quantum Hall effect, with increasing magnetic field, the Hall resistance evolves
from a straight line into step-like behaviors with well-defined plateaux. At the plateaux the Hall conductance
is quantized~\footnote{ In this Section we shall indicate the spatial coordinates as $(x,y)$ instead of $(x_1,x_2)$.}:
\begin{equation}
\label{6.1}
\sigma_{xy}    \;  =   \;  \frac{e^2}{2 \pi \hslash }  \; \nu \;   \;  , \; \;  
\end{equation}
where $\nu$ is  an integer or a certain fraction. At the same time, the longitudinal resistance drops to zero,
suggesting dissipationless transport of charged  quasiparticles. \\
A single two-dimensional Dirac fermion under a magnetic field is known to show the quantized Hall effect with the Hall conductance:
\begin{equation}
\label{6.2}
\sigma_{xy}    \;  =   \;  \frac{e^2}{2 \pi \hslash }  \; (n \; + \; \frac{1}{2}) \; ,
\end{equation}
 $n$ being an integer. The factor  $1/2$ in Eq.~(\ref{6.2}) is  characteristic of
the relativistic dispersion relation of Dirac fermions compared with the usual massive electrons, and it is related to
 the zero modes pinned at the neutral point. Alternatively, the half-integer quantization can be also 
 accounted for as a Berry phase~\cite{Novoselov:2005,Zhang:2005}.
The Hall conductance of three dimensional topological insulators  is expected to be given by the sum
of the  contributions from the top and bottom surfaces (see Fig.~\ref{fig-7}, left panel). Therefore one finds: 
\begin{equation}
\label{6.3}
\sigma_{xy}    \;  =   \;  \frac{e^2}{2 \pi \hslash }  \; (n \; + \; m \; + \; 1) \; ,
\end{equation}
with  $n$ and $m$ are integers referring to the upper and lower surface respectively.
When the  Landau levels of the top and bottom surface coincide exactly, the  
two contributions are equivalent and  one has  $n=m$,  namely only the odd integer quantum Hall effect  is expected.
If, however, the degeneracy of the Landau levels is removed, then  the Hall conductance becomes quantized
as integers. Indeed, recently the integer quantum Hall effect has been observed in three dimensional topological
insulators~\cite{Xu:2014,Yoshimi:2015}. \\
If we admit the generation of a dynamical mass gap, then, according to  Eq.~(\ref{4.5}) we obtain the gapped
Landau level spectrum for the two surfaces as displayed in Fig.~\ref{fig-7}, right panel. 
In fact,  we adopted the convention that magnetic fields perpendicular to the surfaces are positive in the direction of the outward surface normal,
so that  the transverse magnetic field is positive for the top surface and negative for the bottom surface.
It is evident that also in this case we obtain  the odd integer quantum Hall effect for identical surfaces, or the integer
quantum Hall effect if the   the degeneracy of the Landau levels  in the top and bottom surfaces is removed. \\
Interestingly enough, in high magnetic  fields an additional quantum Hall plateau at the charge neutral point  with  $\nu $ = 0 
has been observed~\cite{Yoshimi:2015}. This peculiar zero-Hall plateau
could be explained in a natural way if there are  gapless edge excitations pinned at the Dirac neutral point.
Indeed, it has been already proposed~\cite{Feng:2015} that chiral edge states residing at the magnetic domain boundaries
are responsible for the zero-Hall plateau observed  in the quantum anomalous Hall effect in magnetic topological
insulators. More recently,  chiral edge states residing at the magnetic domain boundaries on the surface
of  a three-dimensional topological insulator have been observed by means of a scanning superconducting quantum 
interference device~\cite{Wang:2015}. \\
In this Section we discuss the presence of current-carrying   chiral edge states residing at the magnetic domain boundaries
between the top and bottom surfaces of a three-dimensional insulator. Moreover, these peculiar edge excitations
are pinned at the Dirac neutral point, whereas the energy spectrum of the surface states is gapped.
It turns out that these  states are localized on the magnetic domain wall and are similar to the Jackiw-Rebbi soliton~\cite{Jackiw:1976}. \\
To face this task, in Fig.~\ref{fig-8} we show a  schematic picture of the magnetic field on the upper and lower surfaces of a three-dimensional
topological insulator. At the edges of the surfaces the magnetic field decreases to zero within a region of linear size given
by the magnetic length:

\begin{equation}
\label{6.4}
a_B  \; = \;  \sqrt{\frac{\hslash c}{e B_0} }   \; .
\end{equation}
The physical meaning of Eq.~(\ref{6.4}) is as follows. The low-lying quantum states localized on the upper surface will feel a magnetic field
which is positive and vanishes near the surface boundaries within a distance of order $\sim a_B$. Obviously, the same happens on the lower surface
where the magnetic field is negative. If we glue together the two surfaces along an edge, as shown in Fig. ~\ref{fig-8}, we see that the surface excitations will be subject to an effective magnetic field which can be written as:
\begin{equation}
\label{6.5}
B(x) \; = \; B_0 \; tgh \left [ \sqrt{\frac{e B_0}{\hslash c}} \;  x \right ]  \; ,
\end{equation}
where $B_0$ is the strength of the (constant) magnetic field inside the two surfaces. Accordingly, the potential vector 
is given by:
\begin{equation}
\label{6.6}
\vec{A}(x,y)  \; = \; \left ( - B(x) \; y \;  \; ,  \; \; 0  \right )  \; \; .
\end{equation}
Let us consider, now, the Dirac equation in presence of the magnetic field given by Eq.~(\ref{6.5}). After taking
into account Eq.~(\ref{6.6}), we readily obtain: 
\begin{equation}
\label{6.7}
 \left ( \begin{array}{cc} 
0   &   \hslash \partial_x - i \frac{e}{c} B(x) y - i \hslash \partial_y \\ 
  - \hslash \partial_x + i \frac{e}{c} B(x)  y - i \hslash \partial_y  &   0 
\end{array}
\right )   \psi (x,y) \; = \; \frac{\varepsilon}{v_F} \psi (x,y) \; ,
\end{equation}
Note that in  Eq.~(\ref{6.7}) we do not consider the mass gap $\Delta_0$. In fact, we are interested here  in 
the zero-energy solutions that are localized at the surface boundary $x=0$ where the magnetic field vanishes. 
Let $\Theta(x,y)$ be the solutions of  Eq.~(\ref{6.7}) with  $\varepsilon=0$. Writing:
\begin{equation}
\label{6.8}
\Theta (x,y) \; = \; 
 \left( \begin{array}{c} 
\theta_1(x,y) \\
\theta_2(x,y)
\end{array} \right)  \; ,
\end{equation}
we get:
\begin{equation}
\label{6.9}
\begin{array}{c}
 \left [ + \hslash  \partial_x  - i \frac{e}{c} B(x)  y - i \hslash \partial_y  \right ]   \theta_2( x,y) \; = \;  0
\\
\; \; \;  \left [  - \hslash  \partial_x  + i \frac{e}{c} B(x)  y - i \hslash \partial_y \right ]   \theta_1(x,y) \; = \;  0   \;  \; .
\end{array} 
\end{equation}  
Let us consider, firstly, the region $x > 0$. For $x \gg a_B$ evidently $B(x) \simeq B_0$. In this case we have already seen that
the solution of    Eq.~(\ref{6.9}) is given by    Eq.~(\ref{3.14}). This suggests to seek the solutions of   Eq.~(\ref{6.9}) in the form:
\begin{equation}
\label{6.10}
\Theta_+ (x,y) \; = \; 
 \left( \begin{array}{c} 
  0 \\
\theta(x,y)
\end{array} \right)  \; \; , \; \; x \; > \; 0 \; \; .
\end{equation}
Inserting   Eq.~(\ref{6.10}) into  Eq.~(\ref{6.9}) we obtain the equation for $\theta(x,y)$ which we rewrite as:
\begin{equation}
\label{6.11}
 \left \{ + \hslash  \partial_x  - i \frac{e}{c} [ B(x) - B_0 ]   y  - i \frac{e}{c} B_0   y
  - i \hslash \partial_y  \right \}   \theta( x,y) \; = \;  0  \; \; .
\end{equation}  
To solve this last equation, we write:
\begin{equation}
\label{6.12}
 \theta( x,y) \; = \;  e^{ - i \frac{e}{\hslash c} y  \int^{+\infty}_{x} [B(x') - B_0] dx' }      
\; \;  \tilde{ \theta} ( x,y)  \; \; ,
\end{equation}  
to get:
\begin{equation}
\label{6.13}
 \left \{ + \hslash  \partial_x  - i \frac{e}{c} B_0   y 
 -  \frac{e}{ c}   \int^{+\infty}_{x} [B(x') - B_0] dx' 
  - i \hslash \partial_y  \right \}  \;  \tilde{\theta}( x,y) \; = \;  0  \; \; .
\end{equation}  
To proceed further, we put:
\begin{equation}
\label{6.14}
 \tilde{\theta}( x,y) \; = \;  \tilde{\theta}_p( x,y)  \; = \;    f(x)  \;  e^{ i p x}  \;  \Phi_{0, p}(y)  \;  \; ,
\end{equation}  
where:
\begin{equation}
\label{6.15}
 \Phi_{0, p}(y)  \;  =   \;  e^{-\frac {1} {2} {\zeta}^2}  \; , \; \zeta \, =\, \sqrt{\frac{eB_0}{\hslash c}} \left( y - \frac{c p}{eB_0} \right )  \;  \; .
\end{equation}
After some manipulations, we finally get:
\begin{equation}
\label{6.16}
\frac{f'(x)}{f(x)}  \;  =   \;   \sqrt{\frac{eB_0}{\hslash c}} \;  g(x  \sqrt{\frac{eB_0}{\hslash c}})  \; \; , \; \; g(z) \; = \; - \, ln \left [ 1 \; + \; e^{ - 2  \, z}  \right ]
\; \; ,
\end{equation}
giving:
\begin{equation}
\label{6.17}
f(x)  \;  =   \; {\cal {N}} \;   e^{\sqrt{\frac{eB_0}{\hslash c}} \;   \int^{x}_{0} g(x'  \sqrt{\frac{eB_0}{\hslash c}})  dx' } \; \; .
\end{equation}
To a  good approximation, for $ x \lesssim  a_B$, we can rewrite  Eq.~(\ref{6.17}) as:
\begin{equation}
\label{6.18}
f(x)  \;  \simeq   \; {\cal {N}} \;   e^{ - \; ln 2 \; \sqrt{\frac{eB_0}{\hslash c}} \; x  } \; \; .
\end{equation}
Putting it all together, we end with:
\begin{equation}
\label{6.19}
\Theta_+ (x,y) \; = \; 
 \left( \begin{array}{c} 
  0 \\
\theta_p(x,y)
\end{array} \right)  \; \; , \; \; x \; > \; 0 \; \; ,
\end{equation}
where:
\begin{equation}
\label{6.20}
 \theta_p( x,y) \; = \;  e^{ - i \frac{e}{\hslash c} y  \int^{+\infty}_{x} [B(x') - B_0] dx' }      
\; \;  \tilde{ \theta}_p( x,y)  \; \; ,
\end{equation}  
and
\begin{equation}
\label{6.21}
 \tilde{\theta}_p( x,y) \; = \;  {\cal {N}} \;   e^{ - \; ln 2 \; \sqrt{\frac{eB_0}{\hslash c}} \; |x|  }   \;  e^{ i p x}  \;  \Phi_{0, p}(y)  \;  \; .
\end{equation}  
It is easy to check that for $x < 0$  the solutions of  Eq.~(\ref{6.9}) are  given by :
\begin{equation}
\label{6.22}
\Theta_- (x,y) \; = \; 
 \left( \begin{array}{c} 
 \theta_{-p}(x,y)  \\
0
\end{array} \right)  \; \; , \; \; x \; < \; 0 \; \; .
\end{equation}
It is worthwhile to stress that the two zero-energy modes $ \Theta_{\pm} (x,y)$ are localized in the region 
$|x| \lesssim   \sqrt{\frac{\hslash c}{e B_0} } $, and are related by parity and time-reversal transformations.
Therefore, if the chemical potential is tuned to the Dirac neutral point, then the zero-energy edge modes
will contribute to the conductivity.  In particular, it is evident that the contributions of these modes cancel
exactly in the Hall conductance $\sigma_{xy} = 0$, while  they add to  the longitudinal conductance 
$\sigma_{xx} = \sigma_{yy}  >  0$. So that, we see that these edge modes give rise to
the  Hall plateau at the charge neutral point  with  $\nu $ = 0, but with a finite longitudinal resistance
at variance of the  usual Hall plateaux where the longitudinal resistance vanishes.
\section{Summary and Conclusions}
\label{S7}
In this work, we have  discussed the  dynamics of low-lying surface excitation
in three-dimensional topological insulators.   We developed a quantum field theoretical description for the
 surface states of three-dimensional topological insulators, which allowed us to investigate 
 the quantum dynamics of low-lying surface states in presence of an applied transverse magnetic field. \\
 We evaluated  the effects of  quantum fluctuations on the ground state. In particular,  we argued that, in
presence of a constant transverse magnetic field, the  quantum fluctuations induce a shift of the energy of 
the Dirac neutral point which, in turns,  depends on the strength of the magnetic field. Moreover,
 we argued that low-lying surface excitations  in three-dimensional topological insulators   
develop a mass gap varying with $\sqrt{eB}$  by  a rearrangement of the Dirac sea
induced by quantum fluctuations. Interestingly enough, very recently the physical consequences of dynamical
mass generation in several classes of topological Dirac metals have been discussed  in
Ref.~\cite{QiSun:2015}.  \\
 To compare our theoretical scenario with observations, we reanalyzed
the available experimental data for the Landau level spectrum of the surface states in three-dimensional
topological insulators.  Remarkably, we argued that  our theoretical results allowed a consistent and coherent
 description of the Landau level spectrum of the surface low-lying excitations. \\ 
 Finally, we showed that recently detected zero-Hall plateau at the charge neutral point
could be accounted for by chiral edge states  residing at the magnetic domain boundaries between the top and bottom surfaces
of three-dimensional topological insulators. 
\appendix
\section{Appendix}
\label{Ap}
We collect here some intermediate steps needed in the derivation of results presented in Sects.~\ref{S2} and \ref{S3}. Firstly,
let us consider the positive and negative energy solutions of the following Dirac equation:
\begin{equation}
\label{A.1}
-i \, \hslash \, v_F \;  \vec {\alpha} \cdot  \vec {\nabla} \;  \psi^{(\pm)}(\vec {x}) \; =  \;  \pm \; \varepsilon_{\vec{p}} \;  \psi^{(\pm)}(\vec {x}) \; ,
\end{equation}
It is quite easy to solve   Eq.~(\ref{A.1}):
\begin{equation}
\label{A.2}
\psi^{(+)}_{\vec{p}}(\vec {x}) \; =  \;  \frac{1}{2 \pi \hslash}  \;       e^{+ i \, \frac{\vec{p} \cdot \vec{x}}{\hslash} } \; u_{\vec{p}}  \; \; , \; \; 
\psi^{(-)}_{\vec{p}}(\vec {x}) \; =  \;  \frac{1}{2 \pi \hslash}  \;       e^{- i \, \frac{\vec{p} \cdot \vec{x}}{\hslash} } \; v_{\vec{p}}  \; \; , \; \; 
 \varepsilon_{\vec{p}} \; = \; v_F \; | \vec{p} | \; ,
\end{equation}
where:
\begin{equation}
\label{A.3}
u_{\vec{p}}  \; = \; \frac{1}{\sqrt{2}}  \left ( 
\begin{array}{c} 
1 \\
- i  \;   e^{+ i \theta_{\vec{p}}} 
\end{array} \right )  \; \; , \; \; 
v_{\vec{p}}  \; = \; \frac{1}{\sqrt{2}}  \left ( 
\begin{array}{c} 
1 \\
+ i  \;   e^{+ i \theta_{\vec{p}}} 
\end{array} \right )  \; \; ,  \; \; 
\tan \theta_{\vec{p}} \; = \; \frac{p_y}{p_x} \; .
\end{equation}
Note that the positive and negative energy solutions of the Dirac equation are normalized according to:
\begin{equation}
\label{A.4}
 \int d^2 x  \;  [ \psi^{(\pm)}_{\vec{p'}}(\vec{x})]^{\dagger} \;  \psi^{(\pm)}_{\vec{p}}(\vec{x})  \; = \;   \delta(\vec{p} - \vec{p'})  \; . 
\end{equation}
Moreover, the Pauli spinors $u_{\vec{p}}$ and $v_{\vec{p}}$ satisfy the following relations:
\begin{equation}
\label{A.5}
u_{\vec{p}} \;  u^{\dagger}_{\vec{p}}  \;  + \;   v_{\vec{p}} \;  v^{\dagger}_{\vec{p}} \; = \; \mathbb{I} \; =
 \left ( \begin{array}{cl} 
1   & 0 \\ 0 & 1 \end{array}
\right )  \; \; ,  \; \;  u^{\dagger}_{\vec{p}} \;  v_{- \vec{p}}  \; = \; 0 \;.
\end{equation}
With the aid of  Eqs.~(\ref{A.4})  and (\ref{A.5}) one can verify that the anticommutation  relations  Eq.~(\ref{2.9})
are indeed equivalent to the canonical     equal time  anticommutation relations  Eq.~(\ref{2.6}). \\
Next,  we are interested in the  Dirac equation in presence of an external magnetic field:
\begin{equation}
\label{A.6}
v_F \; \vec {\alpha} \cdot  \left [  -i   \hslash \vec {\nabla} \; + \;  \frac{e}{c}  \vec { A}(\vec{x}) \right ] 
\psi (\vec {x}) \; = \; \varepsilon \;  \psi (\vec {x}) \; .
\end{equation}
Since we are interested in uniform magnetic fields perpendicular to the crystal surfaces, in the  Landau gauge 
we can write:
\begin{equation}
\label{A.7}
A_k(x_1,x_2) \; = \; -  x_2 \; B  \; {\delta}_{k,1} \;  \; \; \;  k= 1,2 \; .
\end{equation}
As it is well known, in this case  the Dirac equation  Eq.~(\ref{A.6}) can be exactly solved (see, for instance, Ref.~\cite{Akheizer:1965}). 
Indeed,  Eq.~(\ref{A.6}) leads to:
\begin{equation}
\label{A.8}
 \left ( \begin{array}{cc} 
0   &   \hslash \partial_1 - i \frac{eB}{c} x_2 - i \hslash \partial_2 \\ 
  - \hslash \partial_1 + i \frac{eB}{c} x_2 - i \hslash \partial_2  &   0 
\end{array}
\right )   \psi (\vec {x}) \; = \; \frac{\varepsilon}{v_F} \psi (\vec {x}) \; ,
\end{equation}
with $ \partial_i = \frac{\partial}{\partial x_i} , i=1,2$. Writing:
\begin{equation}
\label{A.9}
\psi (\vec {x}) \; = \; e^{ i \frac{ p x_1}{\hslash}} 
 \left( \begin{array}{c} 
\phi_1(x_2) \\
\phi_2(x_2)
\end{array} \right)  \; ,
\end{equation}
we obtain:
\begin{equation}
\label{A.10}
 \begin{array}{c}
\left [ i p  - i \frac{eB}{c} x_2 - i \hslash \partial_2 \right ]   \phi_2(x_2) \; = \;  \frac{\varepsilon}{v_F}     \phi_1(x_2)
 \\
\left [ - i p  + i \frac{eB}{c} x_2 - i \hslash \partial_2 \right ]   \phi_1(x_2) \; = \;  \frac{\varepsilon}{v_F}     \phi_2(x_2) \; .
\end{array} 
\end{equation}  
Inserting the second equation into the first we rewrite  Eq.~(\ref{A.10}) as:  
\begin{equation}
\label{A.11}
\begin{array}{c}
\left [-  \frac{\hslash^2}{2}  \partial_2^2  
 + \frac{1}{2}  \frac{e^2B^2}{c^2} (x_2 - \frac{p c}{eB})^2 \right ]  
 \phi_1(x_2) \; = \;  \frac{1}{2}
 \left [ (\frac{\varepsilon}{v_F})^2 -  \frac{\hslash e B}{c} \right ]    \phi_1(x_2)
 \\
 \phi_2(x_2) \; =  \;  \frac{v_F}{\varepsilon}   \left [ - i p  + i \frac{eB}{c} x_2 - i \hslash \partial_2 \right ]   \phi_1(x_2)  \; . 
\end{array} 
\end{equation}
 The first equation in   Eq.~(\ref{A.11}) is the familiar harmonic oscillator equation. Thus, we can write: 
\begin{equation}
\label{A.12}
 \phi_1(x_2) \; = \;  \Phi_{n,p}(x_2)  \; \; , \; \; 
 \Phi_{n,p}(x_2) \; =  \; N_n \;  H_n(\zeta)  \; e^{-\frac {1} {2} {\zeta}^2} \; ,
\end{equation}  
where: 
\begin{equation}
\label{A.13}
N_n=  \left( \frac {eB} {\pi \hslash c} \right)^{\frac {1}{4}} 
\frac{1}{\sqrt{2^n n!} } \; \; , \; \;
\zeta \, =\, \sqrt{\frac{eB}{\hslash c}} \left( x_2 - \frac{c p}{eB} \right) \; ,
\end{equation}
and $H_n(x)$ being  the Hermite's polynomial of order $n$. Note that our  normalization is such that:
\begin{equation}
\label{A.14}
\int^{+\infty}_{-\infty} d x_2  \; \left | \Phi_{n,p}(x_2) \right |^2 \; = 1\; .
\end{equation}
Moreover, the energy eingenvalues  are:
\begin{equation}
\label{A.15}
\varepsilon \; =\; \pm \, v_F \, \sqrt{ 2 \hslash \frac{eB}{c} (n \, + \, 1)}  \; \; , \; \; n \, = \, 0, \, 1, \, ... 
\end{equation}
One, also, easily find that:
\begin{equation}
\label{A.16}
 \phi_2(x_2) \; = \;  \pm \; \Phi_{n+1,p}(x_2)  \; .
\end{equation}  
Therefore, we are lead to the following  eigenfunctions:
\begin{equation}
\label{A.17}
 \psi^{(\pm)}_{n,p}(x_1,x_2) \;  = \;  \frac {e^{ \pm i p x_1}}{\sqrt {2 \pi \hslash}} \;  \frac{1}{\sqrt{2}} 
 \left( \begin{array}{c} 
\Phi_{n, \pm p}(x_2)  \\
 \pm \, i \, \Phi_{n+1, \pm p}(x_2) 
\end{array} \right) 
\end{equation}
with energy eigenvalues given by Eq.~(\ref{A.15}), respectively.  It is easy to see that there also  zero mode solutions:
\begin{equation}
\label{A.18}
 \psi_{0,p}(x_1,x_2) \; = \;    \frac {e^{ i p x_1}}{\sqrt {2 \pi \hslash}} \; 
 \left( \begin{array}{c} 0 \\    \Phi_{0, p}(x_2) \end{array} \right) \; ,
\end{equation}
with energy eigenvalue:
\begin{equation}
\label{A.19}
\varepsilon_0 \; = \; 0 \; \; . 
\end{equation}
If we adopt the convention that:
\begin{equation}
\label{A.20}
   \Phi_{n, p}(x_2) \; = \; 0 \; \; \; \text{if} \; \; \; n \; < \; 0  \; ,
\end{equation}
then we can write:
\begin{eqnarray}
\label{A.21}
 \psi^{(\pm)}_{n,p}(x_1,x_2) \;  = \;  \frac {e^{ \pm i p x_1}}{\sqrt {2 \pi \hslash}} \;  \frac{1}{\sqrt{2}} 
 \left( \begin{array}{c} 
\Phi_{n-1, \pm p}(x_2)  \\
 \pm \, i \, \Phi_{n, \pm p}(x_2) 
\end{array} \right)  \; \;  \; n \; = \;  1, \; 2,  \;  ....  
\nonumber  \\
 \psi_{0,p}(x_1,x_2) \;  = \;  \frac {e^{ + i p x_1}}{\sqrt {2 \pi \hslash}} \; 
 \left( \begin{array}{c} 
  0  \\
  \Phi_{0,  p}(x_2) 
\end{array} \right)  \; \;  \; n \; = \; 0 \; . 
\end{eqnarray}
with energy eigenvalues:
\begin{equation}
\label{A.22}
\varepsilon_0 \; = \;  0 \; \; , \; \; 
\varepsilon_n^{(\pm)} \; = \; \pm \; \varepsilon_n \; = \; \pm \, v_F \, \sqrt{ 2 \hslash \frac{eB}{c} \, n}  \; \; , \; \; n \, = \, 1, \, 2, \,  ... 
\end{equation}
The wave functions in  Eq.~(\ref{A.21}) are  normalized as:
\begin{equation}
\label{A.23}
\int^{+\infty}_{-\infty} d^2 x \; {\psi^{(\pm)}_{n,p}}^{\dag} 
(\vec {x}) \; \psi^{(\pm)}_{n',p'} (\vec {x})
\; = \; \delta(p-p') \; \delta_{n , n'} \; .
\end{equation}
Note that the Landau levels are infinitely degenerate with density of states:
\begin{equation}
\label{A.24}
\int^{+\infty}_{-\infty} dp \; {\psi^{(\pm)}_{n,p}}^{\dag} 
(\vec {x}) \; \psi^{(\pm)}_{n,p} (\vec {x}) = \frac {eB} {2 \pi \hslash c} \; . 
\end{equation}
To derive Eq.~(\ref{3.11})  we note that by using   Eq.~(\ref{2.20}) we may write:
\begin{equation}
\label{A.25}
 E_0(B) =  - \frac {eB} {2 \pi \hslash c} V  \sum_{n=0}^{\infty}  \varepsilon_{n} \,
= \frac {eB} {2 \pi \hslash c}  \, V \sum_{n=0}^{\infty}   \int_0^{\infty} \frac {ds} {\sqrt{\pi s}} \frac 
{d} {ds} \left(  e^{  - 2 \hslash v_F^2  \frac{eB}{c} \, n \, s}   \right) \; . 
\end{equation}
We may, now, perform the summation over $n$:
\begin{equation}
\label{A.26}
 \sum_{n=0}^{\infty}    e^{  - 2 \hslash v_F^2  \frac{eB}{c} \, n \, s}  \;  = \; 
 \frac{1}{1 -   e^{  - 2 \hslash v_F^2  \frac{eB}{c}  \, s}} \; ,
\end{equation}
to obtain:
\begin{equation}
\label{A.27}
 E_0(B)  \; =  \; 
 \frac {eB} {2 \pi \hslash c}  \, V   \int_0^{\infty} \frac {ds} {\sqrt{\pi s}} \frac 
{d} {ds} \left (     \frac{1}{1 -   e^{  - 2 \hslash v_F^2  \frac{eB}{c}  s}}
  \right ) \; . 
\end{equation}
As discussed in  Sect.~\ref{S2}, the ultraviolet divergences in  $E_0(B)$  are recovered 
from the singularity for  $ s \to  0$. To isolate the divergent part in   Eq.~(\ref{A.27}), we note that:
\begin{equation}
\label{A.28}
 \frac{1}{1 -   e^{  - 2 \hslash v_F^2  \frac{eB}{c}  s}} \;  \stackrel{ s  \to  0}{\sim}  \; 
 + \;  \frac {1} {   2 \hslash v_F^2  \frac{eB}{c}  s    }  \;  \;  \; .
\end{equation}
This suggests  to rewrite   Eq.~(\ref{A.27}) as:
\begin{eqnarray}
\label{A.29}
 E_0(B)  \;  =  && \; 
 \frac {eB} {2 \pi \hslash c}  \, V   \int_0^{\infty} \frac {ds} {\sqrt{\pi s}} \frac 
{d} {ds} \left [    \frac{1}{1 -   e^{  - 2 \hslash v_F^2  \frac{eB}{c}  s}} \;   -  \; \frac {1} {   2 \hslash v_F^2  \frac{eB}{c}  s    }
  \right ] \; 
\nonumber \\
 &&  +  \;  \frac {eB}{2 \pi \hslash c}  \, V   \int_0^{\infty} \frac {ds} {\sqrt{\pi s}} 
 \frac {d}{ds}   \frac{1}{   2 \hslash v_F^2  \frac{eB}{c}  s } 
   \;  . 
\end{eqnarray}
The divergent term in $ E_0(B)$ is due to the second integral in the right-hand part  of  Eq.~(\ref{A.29}). Comparing  with
 Eq.~(\ref{2.21}) we at once recognize that this divergent term coincides with $E_0$. Therefore, we obtain:
\begin{equation}
\label{A.30}
 \Delta E_0(B)  \;  =  \;   E_0(B)  \, - \, E_0 \; = 
 \frac {eB} {2 \pi \hslash c}  \, V   \int_0^{\infty} \frac {ds} {\sqrt{\pi s}} \frac 
{d} {ds} \left [    \frac{1}{1 -   e^{  - 2 \hslash v_F^2  \frac{eB}{c}  s}} \;   -  \; \frac {1} {   2 \hslash v_F^2  \frac{eB}{c}  s    }
   \right ] \; ,
\end{equation}
which agrees with Eq.~(\ref{3.11}).

\end{document}